\documentclass[10pt]{article}

\usepackage{amsmath}
\usepackage{amsfonts}
\makeatletter

\@addtoreset{equation}{section}
\def\rot{\nabla \times}
\def\div{\nabla \cdot}

\def\w{{\omega}}
\def\W{{\Omega}}
\def\vf{{\varphi}}
\def\k{{\kappa}}
\def\l{{\lambda}}
\def\a{{\alpha}}
\def\b{{\beta}}
\def\g{{\gamma}}

\def\e{{\varepsilon}}

\def\triangle{{\Delta}}
\def\G{{\Gamma}}
\def\L{{\Lambda}}
\def\U{{\Upsilon}}

\def\bA{{\bf A}}

\def\bG{{\bf G}}

\def\bg{{\bf g}}
\def\bn{{\bf n}}
\def\bq{{\bf q}}
\def\bv{{\bf v}}
\def\bw{{\bf w}}
\def\bzero{{\bf 0}}

\def\R{{\mathbb R}}

\def\cA{{\mathcal A}}
\def\cB{{\mathcal B}}

\def\cF{{\mathcal F}}
\def\cE{{\mathcal E}}
\def\cH{{\mathcal H}}

\def\cS{{\mathcal S}}
\def\cZ{{\mathcal Z}}
\def\cK{{\mathcal K}}
\def\cL{{\mathcal L}}

\def\bea{\begin{eqnarray*}}
\def\eea{\end{eqnarray*}}

\newtheorem{theor}{Theorem}[section]
\newtheorem{lem}{Lemma}[section]
\newtheorem{prop}{Proposition}[section]
\newtheorem{cor}{Corollary}[section]
\newtheorem{Def}{Definition}[section]
\newtheorem{rem}{Remark}[section]

\def\qed{\hfill\hbox{${\vcenter{\vbox{
\hrule height 0.4pt\hbox{\vrule width 0.4pt height 6pt
\kern5pt\vrule width 0.4pt}\hrule height
0.4pt}}}$}\par\vglue5pt\noindent}

\begin{document}
\title{Global and exponential attractors for a Ginzburg-Landau model of superfluidity}

\author{A. Berti\thanks{Dipartimento di Matematica, Universit\`a di Brescia, 25123 Brescia, Italy, e-mail: alessia.berti@ing.unibs.it}, V. Berti\thanks{Dipartimento di Matematica, Universit\`a di Bologna, 40126 Bologna, Italy, e-mail: berti@dm.unibo.it } $\, $ and I. Bochicchio\thanks{Dipartimento di Matematica e Informatica, Universit\`a di Salerno, 84084 Fisciano (SA), Italy, e-mail: ibochicchio@unisa.it} }
\date{}
\maketitle
\begin{abstract}
The long-time behavior of the solutions for a non-isothermal model in
superfluidity is investigated. The model describes the transition between the normal and the superfluid phase in liquid $^4$He by means of a non-linear differential system, where the concentration of the superfluid phase satisfies a non-isothermal Ginzburg-Landau equation.
This system, which turns out to be consistent with thermodynamical principles and whose well-posedness has been recently proved, has been shown to admit a Lyapunov
functional. This allows to prove existence of the global attractor which consists of the unstable manifold of the stationary
solutions.
Finally,  by exploiting recent techinques of semigroups theory, we prove the existence of an exponential attractor of finite fractal
dimension which contains the global attractor.
\end{abstract}

\noindent
{\bf AMS Classification:} 35B41, 37B25, 82D50.

\bigskip

\noindent
{\bf Keywords:} Superfluids, Ginzburg-Landau equations, Lyapunov functional, exponential attractors.

\section{Introduction}

In this paper we study the asymptotic behavior of the solutions of a Ginzburg-Landau model for superfluidity. This model describes the phase transition between the normal and the superfluid state occurring in liquid helium II when the temperature overcomes a critical value of about $2.2 K$. The phenomenon can be interpreted as a second-order phase transition and accordingly set into the framework of the Ginzburg-Landau theory (see {\it e.g.} \cite{Brokate,F}).
The derivation of this model, its consistency with thermodynamics and the interpretation of some physical aspects related to superfluidity can be found in \cite{F2}. In agreement with Landau's viewpoint, the main matter is to consider each particle of the superfluid as a pair endowed with two different excitations, normal and superfluid, represented respectively by two components $\bv_n$ and $\bv_s$ of the velocity.
The differential system describing the behavior of the superfluid involves three unknowns: the concentration $f$ of the superfluid phase, whose evolution is governed by the Ginzburg-Landau equation, the absolute temperature $u$ which induces the transition and the superfluid component $\bv_s$. The normal component $\bv_n$ is supposed to be expressed in terms of the superfluid velocity through the constitutive equation (see \cite{F2})
$$
\bv_n=\rot\bv_s.
$$
By means of a suitable decomposition of the state variables the differential system ruling the evolution of the superfluid assumes the form
\begin{eqnarray}
\g\psi_{t}&=&\frac{1}{\kappa ^{2}}\triangle \psi-\frac{2i}{\k}\bA\cdot\nabla\psi-\psi|\bA|^2+i\b(\div\bA)\psi\nonumber\\
 \label{P1}
 &&-\psi(|\psi|^{2}-1+u)
 \\
\bA_t&=&\nabla(\div\bA)-\mu\rot\rot\bA- |\psi|^2\bA+\frac{i}{2\k}(\psi\nabla\bar{\psi}-\bar{\psi}\nabla\psi)\nonumber\\
 \label{P2}
&&-\nabla u-\bg\\
c_0 u_t&=&\frac{1}{2}(\psi_t\bar{\psi}+\psi\bar{\psi}_t)+k_0\triangle u\nonumber\\
\label{P3}
&&+
\div\left[-|\psi|^2\bA+\frac{i}{2\k}(\psi\nabla\bar\psi-\bar\psi\nabla\psi)\right]+r
\end{eqnarray}
where $\psi$ is a complex-valued function whose modulus coincides with the concentration of the superfluid phase, and $\bA$ is related to the component $\bv_s$ by
$$\rot\bA=-\rot\bv_s.$$
Equations (\ref{P1})-(\ref{P2}) have the same structure of the Ginzburg-Landau equations of superconductivity (\cite{T}). Indeed, as pointed out by several authors (see {\it e.g.} \cite{Mend, TT}), there are evident analogies between the phenomena of superfluidity and supercondutivity. In this framework, the choice of the decomposition for the unknown variables corresponds to a choice of the gauge for the Ginzburg-Landau equations \cite{BFG, FKT, KT}.

Existence and uniqueness of the global solutions to problem (\ref{P1})-(\ref{P3}) completed with initial and boundary conditions have been proved in \cite{BF}. In this paper we analyze the asymptotic behavior of the solution, by proving first existence of the global attractor and then of exponential attractors.  In the context of superconductivity, the same problem has been treated in \cite{TW}, where the authors prove  existence of the global attractor. Later, Rodriguez-Bernal {\it et al.} \cite{RWW} show that the semigroup generated by the system admits finite-dimensional exponential attractors. The main difference and difficulty in our problem is due to the presence of the absolute temperature which does not appear in the traditional Ginzburg-Landau equations of superconductivity, where an isothermal model is analyzed.
In particular, even if from a physical point of view $u>0$, such a bound cannot be proved a-priori from equations (\ref{P1})-(\ref{P3}). The positivity of the temperature would guarantee the boundedness
\begin{equation}\label{psi<=1}
|\psi|\leq 1,
\end{equation}
 which can be proved in the same way as in superconductivity (\cite{Du}), provided that this inequality holds at the initial instant.
Relation (\ref{psi<=1}) is widely exploited in \cite{BG} and \cite{BFG} to prove that the Ginzburg-Landau system of superconductivity admits absorbing sets, global and exponential attractors.
As a matter of facts the inequality (\ref{psi<=1}) is not used neither in \cite{RWW} nor in \cite{TW}, where existence of the global attractor is proved by means of a Lyapunov functional and exponential attractors are obtained as a consequence of the squeezing property of the solutions (\cite{EFNT}). Therefore, in this paper we construct a Lyapunov functional for system (\ref{P1})-(\ref{P3}) which allows to show existence of the global attractor consisting of the unstable manifold of the stationary solutions.
Furthermore, by means of more recent results devised in \cite{GGMP}, we prove that the semigroup genarated by
(\ref{P1})-(\ref{P3}) possesses an exponential attractor.

The plan of the paper is the following. The model describing the behavior of the superfluid is recalled in section 2. In section 3 we state the existence and uniqueness result obtained in \cite{BF} and prove a-priori estimates and continuous dependence of solutions on the initial data which ensures that the system generates a strongly continuous semigroup on the phase space. Section 4 is devoted to the construction of a Lyapunov functional and to the proof of existence of the global attractor. Finally in section 5, we show that the semigroup admits an exponential attractor.

\section{Statement of the problem}
In this section, we briefly recall the model proposed in \cite{F2} describing the behavior of a superfluid.
Let $\W\subset \R^3$ be the domain occupied by the material. We suppose that $\W$ is bounded with a smooth boundary $\partial \W$, whose unit outward normal will be denoted by $\bn$.
The state variables are identified with the triplet $(f,\bv_s,u)$ representing the concentration of the superfluid phase, the velocity of the superfluid component and the ratio between the absolute temperature and the transition temperature. The evolution of $f$ is ruled by the Ginzburg-Landau equation typical of second order phase transitions (\cite{F}), {\it i.e.}
\begin{equation}
    \label{f}
    \g f_{t}=\frac{1}{\kappa ^{2}}\triangle f-f(f^{2}-1+u+\bv_{s}^{2}),
\end{equation}
where $\g, \k$ are positive constants. The term $\bv_s^2$ allows to prove the existence of a critical velocity above which superfluid properties disappear. Indeed, if $\bv_s$ overcomes a threshold value, the unique solution to (\ref{f}) with boundary and initial conditions
$$
\nabla f\cdot \bn|_{\partial\W}=0\qquad\qquad
f(x,0)=f_0(x)
$$
is $f=0$ that corresponds to the normal phase.

The superfluid component is assumed to solve the equation
\begin{equation}
    \label{vs}
    ({\bv}_{s})_t=-\nabla \phi_{s}- \mu\rot\rot
    \bv_{s}-f^{2}\bv_{s}+\nabla u+\bg,
\end{equation}
where $\mu$ is a positive constant, $\bg$ is a known function related to the body force and $\phi_s$ is a suitable scalar function satisfying
\begin{equation}
\label{divvs}
\div(f^2\bv_s)=-\k^2\g f^2 \phi_s.
\end{equation}

Equations (\ref{vs}) and (\ref{divvs}) are completed by boundary and initial conditions
$$
\bv_{s}\cdot \bn|_{\partial\W}=0,\qquad(\rot\bv_{s})\times \bn|_{\partial\W}={\mbox{\boldmath$\omega$}},
\qquad
\bv_s(x,0)=\bv_{s0}(x).
$$

We notice that (\ref{vs}) and (\ref{divvs}) are similar to equations governing the motion of the superconducting electrons in the framework of superconductivity \cite{T}.
However, in order to account for the thermomechanical effect, the further term $\nabla u$ enters equation (\ref{vs}).
Indeed, since $\nabla u$ has the same sign of the acceleration, an increase of the temperature yields a superfluid flow in the direction of the heat flux.
In this model, we assume that the heat flux $\bq$ satisfies the Fourier constitutive equation
$$
\bq=-k(u)\nabla u,
$$
where the thermal conductivity $k$ depends linearly on the temperature, namely
$$
k(u)=k_0 u,\qquad\qquad k_0>0.
$$
The thermal balance law and the first principle of Thermodynamics lead to the heat equation \cite{F2}
\begin{equation}
\label{u}
c_0 u_t-ff_t=k_0\triangle u+\div(f^2\bv_s)+r,
\end{equation}
where $c_0>0$ is related to the specific heat and $r$ is the heat supply.
The temperature is required to verify the boundary and initial conditions
$$
u|_{\partial\W}=u_{b}\qquad\qquad u(x,0)=u_0(x)\,.
$$
The differential system introduced is proved to be compatible with second law of thermodynamics, since the Clausius-Duhem inequality is satisfied (\cite{BF}).

The functional setting of the differential problem is more convenient if we introduce a suitable decomposition of the variables $\bv_s,\phi_s$, namely
\begin{equation}
\label{vsps}
\bv_s=-\bA+\frac{1}{\k}\nabla\vf,\qquad \phi_s=\div\bA-\frac{1}{\k}\vf_t,
\end{equation}
where $\bA$ and $\vf$ satisfy
\begin{equation}
    \bA\cdot\bn|_{\partial\W}=0,\qquad\qquad \nabla\vf\cdot\bn|_{\partial\W}=0\,.
    \label{A.n}
\end{equation}

In addition, by means of the complex valued function
$$
\psi=f e^{i\vf},
$$
equations (\ref{f})-(\ref{u}) can be written in the form
\begin{eqnarray}
\label{psi0}
\g\psi_{t}&=&\frac{1}{\kappa ^{2}}\triangle \psi-\frac{2i}{\k}\bA\cdot\nabla\psi-\psi|\bA|^2+i\b(\div\bA)\psi\nonumber\\
&&-\psi(|\psi|^{2}-1+u)
 \\
\label{A0}
\bA_t&=&\nabla(\div\bA)-\mu\rot\rot\bA- |\psi|^2\bA+\frac{i}{2\k}(\psi\nabla\bar{\psi}-\bar{\psi}\nabla\psi)\nonumber\\
&&-\nabla u-\bg\\
\label{u0}
c_0 u_t&=&\frac{1}{2}(\psi_t\bar{\psi}+\psi\bar{\psi}_t)+k_0\triangle u\nonumber\\
&&+
\div\left[-|\psi|^2\bA+\frac{i}{2\k}(\psi\nabla\bar\psi-\bar\psi\nabla\psi)\right]+r
\end{eqnarray}
where $\beta=\k\g-{1}/{\k}$ and $\bar\psi$ denotes the complex conjugate of $\psi$.

We associate to (\ref{psi0})-(\ref{u0}) the boundary conditions
\begin{eqnarray*}
\bA\cdot \bn|_{\partial\W}&=&0\qquad(\rot\bA)\times \bn|_{\partial\W}=-{\mbox{\boldmath$\omega$}}\\
\nabla \psi\cdot \bn|_{\partial\W}&=&0\qquad\qquad\qquad \quad
u|_{\partial\W}=u_{b}
\end{eqnarray*}
and initial data
\begin{equation}
\psi(x,0)=\psi_0(x) \qquad \bA(x,0)=\bA_0(x) \qquad u(x,0)=u_0(x).
\label{ic0}
\end{equation}
Furthermore, we assume that $\bg,r,{\mbox{\boldmath$\omega$}},u_b$ are time independent.

\subsection{Notation and functional spaces}
In order to obtain a precise formulation of the problem, we introduce here some notation and recall the main inequalities used in the sequel.

For each $p\geq 1$ and $s\in\R$, we denote by $L^p(\W)$ and $H^s(\W)$ the Lebesgue and Sobolev spaces of real valued, complex valued or vector valued functions, according to the context. Let $\|\cdot\|_p$ and $\|\cdot\|_{H^s}$ be the standard norms of $L^p(\W)$ and $H^s(\W)$, respectively. In particular $\|\cdot\|$ stands for the $L^2(\W)$-norm.
The space $H^1_0(\W)$ is the closure of $C^\infty$ functions with compact support with respect to the norm $\|\cdot\|_{H^1}$.
Finally, we denote by
$$
H^1_{0m} = \left\{ w \in H^1(\W) : \int_\W w \, dv =0 \right\}.
$$

Here and henceforth we denote by $C$ any constant depending only on the domain $\W$ which is allowed to vary even in the same formula. Further dependencies will be specified.

The Sobolev embedding theorem implies (\cite{A})
\begin{eqnarray}
\label{Sob1}
\|w\|_p&\leq& C \|w\|_{H^1},\qquad 1\leq p\leq 6,\quad\ w\in H^1(\W),\\
\label{Sob2}
\|w\|_\infty &\leq& C \|w\|_{H^2},\qquad \qquad \qquad \qquad w\in H^2(\W).
\end{eqnarray}
and the following interpolation inequality holds
\begin{equation}
\label{interp}
\|w\|_3^2\leq C\|w\|\|w\|_{H^1}\qquad w\in H^1(\W).
\end{equation}

If $w\in H^1_0(\W)$ or $w\in H^1_{0m}(\W)$, Poincar\'e inequality provides (\cite{E})
\begin{equation}
\label{poincare}
\|w\|\leq C\|\nabla w\|\,.
\end{equation}
Every $w\in H^2(\W)$ satisfies
\begin{equation}
\label{lapl}
\|w\|_{H^2}\leq C(\|w\|+\|\triangle w\|)\,.
\end{equation}
Furthermore, for every $v\in H^1(\W)$, $w\in H^2(\W)$ the following interpolation inequality holds (\cite{Brokate})
\begin{equation}\label{H1H2}
    \|v w\|_{H^1}\leq C \|v\|_{H^1} \|w\|_{H^2}.
\end{equation}

For vector valued functions we introduce the Hilbert spaces
\begin{eqnarray*}
    \cH^1(\W) &=& \left\{\bw\in H^1(\W):\ \bw\cdot\bn|_{\partial\W}=0\right\}\,,\\
    \cH^2(\W) &=& \left\{\bw\in H^2(\W):\ \bw\cdot\bn|_{\partial\W}=0,\ (\rot\bw)\times\bn|_{\partial \W}=\bzero\right\}\,.
\end{eqnarray*}

\begin{lem}
The spaces $\cH^1(\W), \cH^2(\W)$ are Hilbert spaces with respect to  the norms
\begin{eqnarray}
\label{H1norm}
\|\bw\|_{\cH^1}^2&=&\|\div\bw\|^2+\|\rot\bw\|^2,
\\
\label{H2norm}
\|\bw\|_{\cH^2}^2&=&\|\nabla(\div\bw)\|^2+\|\rot\rot\bw\|^2.
\end{eqnarray}
In particular, the following estimates hold
\begin{eqnarray}
\label{poincv}
C_1\|\bw\|^2_{\cH^1}\leq \|\bw\|_{H^1}^2\leq C_2\|\bw\|^2_{\cH^1},
\qquad\qquad \bw\in\cH^1(\W)
\\ \label{poincv2}
C_3\|\bw\|^2_{\cH^2} \leq \|\bw\|_{H^2}^2\leq C_4 \|\bw\|^2_{\cH^2},
\qquad\qquad \bw\in\cH^2(\W)
\end{eqnarray}

\end{lem}

\noindent
{\bf Proof.}
The inequalities (\ref{poincv}) follow from \cite[Prop. 3.2]{Nibbi}.

Let $\bw\in \cH^2(\W)$. The identity
$$
\Delta \bw = \nabla (\div\bw) - \rot\rot\bw
$$
and (\ref{lapl})  yield
\begin{equation}\label{interm}
\|\bw\|_{H^2}^2\leq C(\|\bw\|^2+\|\nabla(\div\bw)\|^2+\|\rot\rot\bw\|^2).
\end{equation}
Moreover, by means of (\ref{Sob1}), (\ref{H1norm}) and (\ref{poincv}), we obtain
$$
\|\bw \|^2 \leq \|\bw\|^2_{H^1} \leq C_2\left(\|\div\bw\|^2 + \|\rot\bw\|^2  \right).
$$
The boundary condition $\bw\cdot\bn| _{\partial\W}=0$ ensures that $\div\bw \in H^1_{0m}(\W)$.
Hence, (\ref{poincare}) yields
$$
\|\div\bw\| \leq  C\|\nabla(\div\bw)\|.
$$
Finally, by applying (\ref{H1norm}) and \cite[Prop. 2.2]{Nibbi}, we prove
$$
\|\rot\bw\| \leq \|\rot\bw\|_{H^1} \leq C\|\rot\rot\bw\|.
$$

Substitution into (\ref{interm}) leads to (\ref{poincv2}).
\qed

\section{Well-posedness of the problem}
\subsection{Existence and uniqueness}
In order to deal with homogeneous boundary conditions, we consider the new variables
$$
\hat\bA=\bA-\bA_\cH,\qquad\hat u=u-u_\cH,
$$
where $\bA_\cH$ and $u_\cH$ are solutions of the problems
\begin{eqnarray*}
\left\{
\begin{array}{lll}
\rot\rot\bA_\cH=0\\ \div\bA_\cH=0\\
\bA_\cH\cdot \bn|_{\partial\W}  =  0\\ (\rot\bA_\cH)\times  \bn|_{\partial\W}=-{\mbox{\boldmath$\omega$}}\\
\end{array}
\right.\qquad\qquad
\left\{
\begin{array}{ll}
\triangle u_\cH=0\\
u_\cH|_{\partial    \W}=u_b
\end{array}
\right.
\end{eqnarray*}
From the standard theory of linear partial differential equations, it follows that if
${\mbox{\boldmath$\omega$}}\in H^{1/2}(\partial\W)$, $u_b\in H^{1/2}(\partial\W)$, then
$$
\bA_\cH\in \cH^1(\W),\qquad\qquad u_\cH\in H^1(\W)
$$
and
$$
\|\bA_\cH\|_{ \cH^1}\leq C \|{\mbox{\boldmath$\omega$}}\|_{ H^{1/2}(\partial\W)},\qquad\qquad
\|u_\cH\|_{ H^1}\leq C\|u_b\|_{ H^{1/2}(\partial\W)}.
$$
Accordingly, system (\ref{psi0})-(\ref{ic0}) can be written as
\begin{eqnarray}
\label{psi_i}
\g\psi_{t}&=&\frac{1}{\kappa ^{2}}\triangle \psi-\frac{2i}{\k}(\hat\bA+\bA_\cH)\cdot\nabla\psi-\psi|\hat\bA+\bA_\cH|^2+i\b(\div\hat\bA)\psi
\nonumber\\
&&-\psi(|\psi|^{2}-1+\hat u + u_\cH)
 \\
 \label{A_i}
\hat\bA_t&=&\nabla(\div\hat\bA)-\mu\rot\rot\hat\bA- |\psi|^2(\hat\bA+\bA_\cH)+\frac{i}{2\k}(\psi\nabla\bar{\psi}-\bar{\psi}\nabla\psi)
\nonumber\\
&&-\nabla \hat u- \nabla u_\cH -\bg\\
\label{u_i}
c_0 \hat u_t&=&\frac{1}{2}(\psi_t\bar{\psi}+\psi\bar{\psi}_t)+k_0\triangle \hat u
\nonumber\\
&&+
\div\left[-|\psi|^2(\hat\bA+\bA_\cH)+\frac{i}{2\k}(\psi\nabla\bar\psi-\bar\psi\nabla\psi)\right]+r
\end{eqnarray}
with boundary conditions
\begin{eqnarray}
\label{bc1_i}
\hat\bA\cdot \bn|_{\partial\W}&=&0\qquad(\rot\hat\bA)\times \bn|_{\partial\W}={\bf 0}
\\
\label{bc2_i}
\nabla \psi\cdot \bn|_{\partial\W}&=&0\qquad\qquad\qquad \quad
\hat u|_{\partial\W}=0
\end{eqnarray}
and initial data
\begin{equation}
\psi(x,0)=\psi_0(x) \qquad \hat \bA(x,0)=\hat\bA_0(x) \qquad \hat u(x,0)=\hat u_0(x),
\label{ic_i}
\end{equation}
where $\hat\bA_0(x)= \bA_0(x) - \bA_{\cH}(x)$ and $\hat u_0(x)= u_0(x) - u_{\cH}(x)$.

We denote by $z=(\psi,\hat\bA,\hat u)$ and introduce the functional spaces
\begin{eqnarray*}
\cZ^1(\W)&=&H^1(\W)\times\cH^1(\W)\times L^2(\W),\\
\cZ^2(\W)&=&H^2(\W)\times\cH^2(\W)\times H^1_0(\W),
\end{eqnarray*}
endowed respectively with the norms
\begin{eqnarray*}
\|z(t)\|_{\cZ^1}&=&(\|\psi(t)\|^2_{H^1}+\|\hat\bA(t)\|^2_{\cH^1}+\|\hat u(t)\|^2)^{1/2}\\
\|z(t)\|_{\cZ^2}&=&(\|\psi(t)\|^2_{H^2}+\|\hat\bA(t)\|^2_{\cH^2}+\|\hat
u(t)\|^2_{H^1_0})^{1/2}.
\end{eqnarray*}
Existence and uniqueness of solutions to problem (\ref{psi_i})-(\ref{ic_i}) have been shown in \cite{BF}.
For convenience we recall this result.

\begin{theor} \label{exist} Let  $z_0 = (\psi_0,\hat\bA_0,\hat u_0)\in \cZ^1(\W)$, $\bA_\cH\in\cH^1(\W)$, $u_\cH\in H^1(\W)$, $\bg,r\in L^2(\W)$. Then, for every $T>0$, there exists a unique solution $z$ of the problem (\ref{psi})-(\ref{ic}) such that
\begin{eqnarray*}
\psi & \in &  L^2(0,T,H^1(\W))\cap H^1(0,T, L^2(\W)) \\
\hat\bA & \in &  L^2(0,T,\cH^1(\W))\cap H^1(0,T, L^2(\W))\\
\hat u &\in & L^2(0,T,H^1_0(\W))\cap H^1(0,T,H^{-1}(\W)).
\end{eqnarray*}
Moreover $\psi\in L^2(0,T,H^2(\W))\,\cap\, C(0,T,H^1(\W))$,
${\hat\bA}\in  L^2(0,T,$ $\cH^2(\W))\,\cap\, C(0,T,$ $\cH^1(\W))$,
$\hat{u}\in$ $C(0,T,L^2(\W))$.
\end{theor}

\subsection{A-priori estimates}

Henceforth we assume that $\bg\in H^1(\W)$ and
$$
\div\bg =0, \qquad\qquad r=0.
$$
In particular, since $\triangle u_\cH=0$, there exists a vector-valued function $\bG$ such that
$$
\nabla u_\cH + \bg = \rot\bG.
$$
Moreover, $\bG$ is defined to within the gradient of an arbitrary scalar function. Therefore it is not restrictive to assume the boundary condition
\begin{equation}\label{bcG}
\bG\times\bn|_{\partial\W}={\bf 0}.
\end{equation}
With these assumptions, system (\ref{psi_i})-(\ref{ic_i}) reduces to
\begin{eqnarray}
\label{psi}
\g\psi_{t}&=&\frac{1}{\kappa ^{2}}\triangle \psi-\frac{2i}{\k}(\hat\bA+\bA_\cH)\cdot\nabla\psi-\psi|\hat\bA+\bA_\cH|^2+i\b(\div\hat\bA)\psi
\nonumber\\
&&-\psi(|\psi|^{2}-1+\hat u + u_\cH)
 \\
 \label{A}
\hat\bA_t&=&\nabla(\div\hat\bA)-\mu\rot\rot\hat\bA- |\psi|^2(\hat\bA+\bA_\cH)+\frac{i}{2\k}(\psi\nabla\bar{\psi}-\bar{\psi}\nabla\psi)
\nonumber
\\
&&-\nabla \hat u- \rot\bG
\\
\label{u2}
c_0 \hat u_t&=&\frac{1}{2}(\psi_t\bar{\psi}+\psi\bar{\psi}_t)+k_0\triangle \hat u
\nonumber\\
&&+
\div\left[-|\psi|^2(\hat\bA+\bA_\cH)+\frac{i}{2\k}(\psi\nabla\bar\psi-\bar\psi\nabla\psi)\right]
\end{eqnarray}
with boundary conditions
\begin{eqnarray}
\label{bc1}
\hat\bA\cdot \bn|_{\partial\W}&=&0\qquad(\rot\hat\bA)\times \bn|_{\partial\W}={\bf 0}
\\
\label{bc2}
\nabla \psi\cdot \bn|_{\partial\W}&=&0\qquad\qquad\qquad \quad
\hat u|_{\partial\W}=0
\end{eqnarray}
and initial data
\begin{equation}\label{ic}
\psi(x,0)=\psi_0(x) \qquad \hat \bA(x,0)=\hat\bA_0(x) \qquad \hat u(x,0)=\hat u_0(x).
\end{equation}

\begin{prop}\label{a-priori}
The solution of (\ref{psi})-(\ref{ic}) with initial datum $z_0\in \cZ^1(\W)$ such that $\|z_0\|_{\cZ^1}\leq R$, satisfies the following a-priori estimates
\begin{eqnarray}
\label{apriori_z1}
&&\sup_{t\geq 0}(\|\psi(t)\|_{H^1}+\|\hat\bA(t)\|_{\cH^1}+\|\hat u(t)\|)\leq C_R,
\\
\label{apriori_z2}
&&\sup_{t\geq 0}\int_0^t\left(\|\psi_t\|^2+\|\hat\bA_t\|^2\right)ds \leq C_R,
\\
\label{apriori_z3}
&&\int_0^t\left(\|\psi\|_{H^2}^2+\|\hat\bA\|_{\cH^2}^2+\|\hat u\|_{H^1_0}^2\right)ds \leq C_R(1+t), \qquad\ t>0,
\\
\label{apriori_z4}
&&\int_t^{t+1}\left(\|\psi\|_{H^2}^2+\|\hat\bA\|_{\cH^2}^2+\|\hat u\|_{H^1_0}^2\right)ds \leq C_R, \qquad\qquad t>0,
\end{eqnarray}
where $C_R$ depends increasingly on $R$.
\end{prop}

\noindent
{\bf Proof.}
Let
\begin{eqnarray}\nonumber
    \cL(\psi,\hat\bA,\hat u) &=& \frac12\int_\W \left\{\left|\frac{i}{\k}\nabla\psi + \psi\hat\bA +\psi \bA_\cH\right|^2 + \frac{1}{2}(|\psi|^2-1)^2 +  |\psi|^2u_\cH   \right.
    \\ \label{Lyap}
    &&+ \mu|\rot\hat\bA|^2+  \eta(\div\hat\bA)^2 +2\rot\bG \cdot \hat\bA + c_0\hat u^2 \Big\} dv,
\end{eqnarray}
where $\eta=2k_0/(k_0+1)$.

Firstly, we show that $\cL$ is non-increasing. By differentiating (\ref{Lyap}) with respect to $t$, we obtain
\bea
\frac{d\cL}{dt} &=&
\int_\W\bigg\{ \frac1{2\k^2}(\nabla\psi_t \cdot \nabla\bar\psi + \nabla\bar\psi_t \cdot \nabla\psi )
 + \frac{i}{2\k} (\bar\psi\nabla\psi_t - \psi \nabla\bar\psi_t) \cdot (\hat\bA+\bA_\cH)
\\&&
- \frac{i}{2\k} (\psi_t \nabla\bar\psi -\bar\psi_t\nabla\psi ) \cdot (\hat\bA+\bA_\cH)
-\frac{i}{2\k} (\psi\nabla\bar\psi - \bar\psi \nabla\psi) \cdot \hat\bA_t
\\ &&
+
\frac{1}{2} (\psi_t\bar\psi + \bar\psi_t\psi) |\hat\bA+\bA_\cH|^2+ |\psi|^2\hat\bA_t \cdot (\hat\bA+\bA_\cH)
\\&&
 +\frac12 (|\psi|^2-1+u_\cH) (\psi\bar\psi_t + \bar\psi\psi_t) + \mu\rot\hat\bA \cdot \rot\hat\bA_t 
\\&&
     + \eta(\div\hat\bA)( \div\hat\bA_t) + \rot\bG \cdot \hat\bA_t + c_0\hat u \hat u_t \bigg\} dv
\eea
By integrating by parts and using (\ref{bc1})-(\ref{bc2}), the terms in the previous expression can be written in the form
\bea
\frac{d\cL}{dt} &=&
\int_\W\bigg\{ \frac12\psi_t\left[-\frac1{\k^2}\triangle\bar\psi - \frac{2i}{\k} \nabla\bar\psi\cdot (\hat\bA+\bA_\cH) +
\bar\psi|\hat\bA+\bA_\cH|^2\right.\\
&& + (|\psi|^2-1+u_\cH) \bar\psi\bigg]+\frac{1}{2}\bar\psi_t\bigg[-\frac1{\k^2}\triangle\psi + \frac{2i}{\k} \nabla\psi\cdot (\hat\bA+\bA_\cH) 
\\&&
+
\psi|\hat\bA+\bA_\cH|^2
+(|\psi|^2-1+u_\cH)\psi\bigg]+\hat\bA_t\cdot\bigg[-\frac{i}{2\k} (\psi\nabla\bar\psi - \bar\psi \nabla\psi)
\\ &&
 +|\psi|^2 (\hat\bA+\bA_\cH)+\mu\rot\rot\hat\bA+\rot\bG\bigg]
\\&&
- \frac{i}{2\k}(\psi_t\bar\psi-\bar\psi_t\psi)\div\hat\bA  + \eta(\div\hat\bA)(\div\hat\bA_t)+ c_0\hat u \hat u_t \bigg\} dv.
\eea
By substituing (\ref{psi})-(\ref{u2}), we obtain
\begin{eqnarray}
\frac{d\cL}{dt} &=&
\int_\W\left\{ -\g|\psi_t|^2 - \frac{i}{2\k}(1+\k\b)(\psi_t\bar\psi-\bar\psi_t\psi)\div\hat\bA
-|\hat\bA_t|^2 \right.\nonumber\\
&&+\hat\bA_t\cdot\left[\nabla(\div\hat\bA) -\nabla\hat u\right] - k_0|\nabla\hat u|^2+ \eta(\div\hat\bA) (\div\hat\bA_t)
\nonumber
\\&&
\left. -\nabla\hat u \cdot \left[ -|\psi|^2(\hat\bA + \bA_\cH) + \frac{i}{2\k}(\psi\nabla\bar\psi-\bar\psi\nabla\psi)\right] 
\right\} dv\nonumber
\\ &=& \int_\W\bigg\{ -\g|\psi_t|^2 - \frac{i\k\g}{2}(\psi_t\bar\psi-\bar\psi_t\psi)\div\hat\bA
-|\hat\bA_t|^2  - k_0|\nabla\hat u|^2
\nonumber
\\&&
\label{ly}
\quad  -\nabla\hat u \cdot \left[ 2\hat\bA_t - \nabla(\div\hat\bA) + \nabla\hat u \right] -
(\eta-1)\hat\bA_t \cdot \nabla(\div\hat\bA)
\bigg\} dv,\nonumber\\
\end{eqnarray}
since $\beta=\k\g-{1}/{\k}$.

We let
\bea
q(\hat\bA_t,\nabla(\div\hat\bA),\nabla\hat u)&=&|\hat\bA_t|^2+|\nabla(\div\hat\bA)|^2+ (k_0+1)|\nabla\hat u|^2\\
&+&(\eta-2)\hat\bA_t\cdot\nabla(\div\hat\bA)+
2\hat\bA_t\cdot\nabla\hat u-2\nabla(\div\hat\bA)\cdot\nabla\hat u.
\eea
A direct check proves that $q$ is a positive definite quadratic form, since
$\eta=2k_0/(k_0+1)$.

Owing to the identity
$$
|\psi_t|^2=|\psi_t-i\k\psi\div\hat\bA|^2-\k^2|\psi|^2(\div\hat\bA)^2-i\k(\psi_t \bar\psi-\bar\psi_t\psi)\div\hat\bA,
$$
equation (\ref{ly}) reads
\begin{eqnarray*}
\frac{d\cL}{dt} =
 \int_\W\Big\{ -\g|\psi_t-i\k\psi\div\hat\bA|^2 + \frac{i\k\g}{2}(\psi_t\bar\psi-\bar\psi_t\psi)\div\hat\bA+\k^2\g |\psi|^2(\div\hat\bA)^2
\nonumber
\\
  -q(\hat\bA_t,\nabla(\div\hat\bA),\nabla\hat u)-\nabla\hat u \cdot  \nabla(\div\hat\bA)  -
\hat\bA_t \cdot \nabla(\div\hat\bA)+|\nabla(\div\hat\bA)|^2
\Big\} dv.
\end{eqnarray*}
Taking (\ref{psi}) into account, we obtain
\begin{eqnarray*}
\frac{d\cL}{dt} &=&
 \int_\W\Big\{ -\g|\psi_t-i\k\psi\div\hat\bA|^2 + \frac{i\k}{2}\Big[\frac{1}{\k^2}(\bar\psi\triangle\psi-\psi\triangle\bar\psi)\nonumber\\
 &&
 \qquad-\frac{2i}{\k}(\hat\bA+\bA_\cH)\cdot(\bar\psi\nabla\psi+\psi\nabla\bar\psi) +2i\b|\psi|^2\div\hat\bA\Big]\div\hat\bA
\\&&
\qquad  +\k^2\g |\psi|^2(\div\hat\bA)^2-q(\hat\bA_t,\nabla(\div\hat\bA),\nabla\hat u)
\nonumber\\
 &&
 \qquad-[\nabla\hat u   +
\hat\bA_t -\nabla(\div\hat\bA)]\cdot  \nabla(\div\hat\bA)
\Big\} dv
\end{eqnarray*}
Furthermore, in view of (\ref{A}), the previous equation can be reduced to
\begin{eqnarray*}
\frac{d\cL}{dt}&=&
\int_\W\Big\{ -\g|\psi_t-i\k\psi\div\hat\bA|^2 + \Big[\frac{i}{2\k}(\bar\psi\triangle\psi-\psi\triangle\bar\psi)\nonumber\\
 &&
 \qquad+(\hat\bA+\bA_\cH)\cdot(\bar\psi\nabla\psi+\psi\nabla\bar\psi) +|\psi|^2\div\hat\bA\Big]\div\hat\bA
\nonumber
\\&&
\qquad  -q(\hat\bA_t,\nabla(\div\hat\bA),\nabla\hat u)\nonumber
\\&&
\qquad
+\Big[|\psi|^2(\hat\bA+\bA_\cH)-\frac{i}{2\k}(\psi\nabla\bar\psi-\bar\psi\nabla\psi)\Big]\cdot  \nabla(\div\hat\bA)
\Big\} dv.
\end{eqnarray*}
The terms involving $\rot\rot\hat\bA$ and $\rot\bG$ vanish by means of an integration by parts owing to (\ref{bc1}) and (\ref{bcG}).

Finally, a further integration by parts leads to
\begin{equation}\label{Lq}
    \frac{d\cL}{dt} = \int_\W\left[ -\g|\psi_t-i\k\psi\div\hat\bA|^2-q(\hat\bA_t,\nabla(\div\hat\bA),\nabla\hat u)\right]dv\leq 0.
\end{equation}
Accordingly, $\cL$ is non-increasing.

We define
\bea
\cF_1(\psi,\hat\bA,\hat u) &=&  \left\|\frac{i}{\k}\nabla\psi + \psi\hat\bA +\psi \bA_\cH\right\|^2 + \frac{1}{2}\|\,|\psi|^2-1\,\|^2      \\
 &&+  \int_\W |\psi|^2u_\cH dv+ \|\hat\bA\|_{\cH_1}^2 + \|\hat u\|^2.
\eea
An application of H\"older's and Young's inequality leads to
\begin{equation}
\label{equiv}
c_1 \cF_1 - c_2 \leq \cL \leq c_3\cF_1 + c_2,
\end{equation}
where $c_1, c_2, c_3$ are suitable positive constants.

Moreover,
\bea
\left\|\frac{i}{\k}\nabla\psi\right\| &\leq& \left\|\frac{i}{\k}\nabla\psi + \psi\hat\bA +\psi \bA_\cH\right\| + \|(\hat\bA+\bA_\cH)\psi\|
     \\
     &\leq &
     \left\|\frac{i}{\k}\nabla\psi + \psi\hat\bA +\psi \bA_\cH\right\| +C \left(\|\hat\bA\|_{6}+\|\bA_\cH\|_{6}\right)\|\psi\|_3,
\eea
so that, by means of (\ref{Sob1}), (\ref{interp}) and Young's inequality, we obtain
\bea
\frac{1}{\k}\left\|\nabla\psi\right\|&\leq& \left\|\frac{i}{\k}\nabla\psi + \psi\hat\bA +\psi \bA_\cH\right\|+ C\left(\|\hat\bA\|_{\cH^1}^2+\|\bA_\cH\|_{\cH^1}^2+1\right)\|\psi\|\\
&& +\frac{1}{2\k}\left\|\nabla\psi\right\|,
\eea
which leads to the estimate
\begin{equation}\label{Egradpsi}
\left\|\nabla\psi\right\|^2\leq C\left[\left\|\frac{i}{\k}\nabla\psi + \psi\hat\bA +\psi \bA_\cH\right\|^2+(\|\hat\bA\|_{\cH^1}^4+1)\|\psi\|^2 \right].
\end{equation}
In addition, H\"older's inequality yields
\begin{equation}\label{Epsi}
\|\psi\|^2 \leq C\|\psi^2\| = C(\|\, |\psi|^2 -1\,\| + 1).
\end{equation}
From the definition of $\cF_1$ and relations (\ref{Egradpsi})-(\ref{Epsi}) we deduce
\begin{equation}\label{zF1}
    \|z(t)\|_{\cZ^1}^2\leq C\left[1+\cF_1(z(t))+\cF_1^2(z(t))+\cF_1^3(z(t))\right].
\end{equation}
Since $\cL(z(t)) \leq \cL(z(0))$, (\ref{equiv}) and (\ref{zF1}) yield (\ref{apriori_z1}).

By integrating (\ref{Lq}) with respect to $t$ we obtain
\begin{eqnarray}
\label{apriori_z5}
    &&\int_0^t [\|\hat\bA_t \|^2 + \|\nabla(\div\hat\bA) \|^2 + \|\nabla\hat u \|^2]ds \leq C_R,
    \\
    &&\int_0^t \|\psi_t-i\k\psi\div\hat\bA\|^2 ds \leq C_R.
\end{eqnarray}
In view of H\"older's inequality and the Sobolev embedding theorem, we have
\bea
\int_0^t \|\psi_t\|^2 ds &\leq& 2\int_0^t [\|\psi_t -i\k \psi\div\hat\bA\|^2 + \k^2 \|\psi\div\hat\bA\|^2]ds\\
& \leq&
C_R + C\int_0^t \|\psi\|^2_{H^1}\|\nabla(\div\hat\bA)\|^2 ds \leq C_R
\eea
where the last inequality follows from (\ref{apriori_z1}) and (\ref{apriori_z5}). Hence (\ref{apriori_z2}) holds.

Finally, (\ref{apriori_z1}), (\ref{apriori_z2}) and a comparison with (\ref{psi})-(\ref{A}) lead to (\ref{apriori_z3}).
By repeating the same arguments, one can easily prove (\ref{apriori_z4}).

\qed

\subsection{Continuous dependence}
The following theorem proves the continuous dependence of the solutions to (\ref{psi})-(\ref{ic}) on the initial data.
\begin{theor}\label{cont_dep}
Let $z_i=(\psi_i,\hat\bA_i,\hat u_i)$, $i=1,2$ be two solutions of (\ref{psi})-(\ref{ic}) with data
$(\bA_{\cH},u_{\cH},\bG)\in\cH^1(\W)\times H^1(\W)\times H^1(\W)$ and $z_{0i}=(\psi_{0i},\bA_{0i},u_{0i})\in \cZ^1(\W)$, $i=1,2$. Then, there exists a constant $C_R$ such that
\begin{eqnarray*}
&&\|z_1(t) - z_2(t)\|_{\cZ^1}^2
 \leq C_R e^{C_R t}\, \| z_{01} -z_{02}\|^2_{\cZ^1} .
\end{eqnarray*}
Moreover, inequality
\begin{equation}\label{int_Z2}
\int_0^t \|z_1(t) - z_2(t)\|_{\cZ^2}^2 \, d s \leq C(t) \| z_{01} -z_{02}\|^2_{\cZ^1}
\end{equation}
holds, where $C(t)$ is a suitable function depending on $t$.
\end{theor}

\noindent {\bf Proof.}
We denote by $\psi=\psi_1-\psi_2$,
$\hat\bA=\hat{\bA}_1-\hat{\bA}_2$, $\hat u=\hat{u}_1-\hat{u}_2$. Equations (\ref{psi})-(\ref{u2}) lead to
\begin{eqnarray}
\label{psid}
  &&\gamma\psi_t-\frac 1{\kappa^2} \triangle \psi+\frac {2i}{\kappa} [\hat\bA\cdot \nabla \psi_1+(\hat\bA_2+\bA_{\cH})\cdot \nabla
  \psi]+|\hat\bA_1+\bA_{\cH}|^2 \psi+
  \nonumber\\
  &&\  \  \quad+ \psi_2(\hat\bA_1 + \hat\bA_2 + 2\bA_{\cH})\cdot \hat\bA -i\b(\psi\div\hat\bA_1+\psi_2\div\hat\bA)-\psi
  \nonumber\\
  &&\  \ \quad+\psi|\psi_1|^2 +\psi_2(\bar\psi_1\psi+\psi_2\bar\psi) +\psi
  (\hat{u}_1+u_{\cH})+\psi_2\hat{u}=0\\[0.7em]
\label{Ad}
  && \hat{\bA}_t-\nabla(\div \hat{\bA})+\mu\rot\rot\hat\bA+|\psi_1|^2
  \hat{\bA}+(\bar\psi_1\psi+\psi_2\bar\psi)(\hat\bA_2+\bA_{\cH})
  \nonumber\\
  &&\ \ \quad-\frac {i}{2\kappa} \big(\psi\nabla \bar\psi_1-\bar \psi \nabla \psi_1+\psi_2\nabla \bar\psi-\bar \psi_2 \nabla \psi\big)
  +\nabla \hat u ={\bf 0}
  \\[0.7em]
\label{ud}
  && c_0 \hat{u}_t-k_0\triangle {\hat u}-\frac12 \big(\psi \bar\psi_{1t}+\bar \psi \psi_{1t}+\psi_2 \bar\psi_{t}+\bar \psi_2 \psi_{t}\big)
  -\div\Big[-|\psi_1|^2 \hat{\bA}
  \nonumber\\
  &&\quad-(\bar\psi_1\psi+\psi_2\bar\psi) (\hat {\bA}_2+\bA_{\cH})+
  \frac {i}{2\kappa} \big(\psi\nabla \bar\psi_1-\bar \psi \nabla \psi_1+\psi_2\nabla \bar\psi-\bar \psi_2 \nabla \psi\big)\Big] =0
  \nonumber\\
\end{eqnarray}
Let us multiply  (\ref{psid}) by $1/2\,(\bar{\psi}+\bar{\psi}_t)$, its
conjugate by $1/2\,(\psi+\psi_t)$, (\ref{Ad}) by $\hat \bA_t$, (\ref{ud}) by $\hat u$ and add the
resulting equations. An integration over $\W$ yields the equality
\begin{eqnarray}\label{d}
&&\frac{1}{2}\frac{d}{dt}\Big[\gamma\|\psi\|^2+\frac{1}{\kappa^2}\|\nabla\psi\|^2+\|\div \hat\bA\|^2+\mu\|\rot\hat \bA\|^2 +c_0\|\hat{u}\|^2\Big]+ \frac{1}{\kappa^2}\|\nabla\psi\|^2\nonumber\\
&&+\gamma\|\psi_t\|^2+ \|\hat
\bA_t\|^2
 +k_0\|\nabla\hat{u}\|^2 +\int_\W\left[|\hat\bA_1+\bA_{\cH}|^2|\psi|^2+|\psi_1|^2|\psi|^2\right]dv\nonumber\\
&&
=\|\psi\|^2-\sum_{h=1}^8 I_h\,,
\end{eqnarray}
where
\begin{eqnarray*}
I_1&=& \frac{i}{\kappa}\int_\W\left[\hat \bA  \cdot(\bar{\psi}\nabla\psi_1 - \psi\nabla\bar{\psi}_1)+
(\hat \bA_2 + \bA_{\cH})\cdot(\bar{\psi}\nabla\psi-\psi\nabla\bar{\psi})\right]dv\nonumber\\
I_2&=&\frac{1}{2}\int_\W(\psi_2\bar{\psi}+\bar{\psi}_2\psi)[(\hat \bA_1+ \hat \bA_2+2\bA_{\cH})\cdot \hat \bA +\bar\psi_1{\psi}+{\psi}_2\bar\psi+\hat {u} ]dv\nonumber\\
I_3&=&-\frac{i\b}{2}\int_\W\left[(\psi\bar\psi_t-\bar\psi\psi_t)\div \hat \bA_1+(\psi_2\bar\psi-\bar\psi_2\psi+\psi_2\bar\psi_t-
\bar\psi_2\psi_t)\div \hat \bA\right]dv\nonumber\\
I_4&=&\int_\W \left\{(\hat u_1 + u_{\cH})|\psi|^2+\frac{1}{2}(\psi\bar{\psi}_t+\bar{\psi}\psi_t)\left[|\hat \bA_1+\bA_{\cH}|^2-1+|\psi_1|^2+\hat u_1+u_{\cH}\right]\right\}dv\nonumber\\
I_5&=& \frac{i}{\kappa}\int_\W\left[\hat \bA \cdot( \bar{\psi}_t\nabla\psi_1
-\psi_t\nabla \bar{\psi}_1)+(\hat \bA_2+\bA_{\cH})\cdot(\bar{\psi}_t\nabla\psi - \psi_t\nabla \bar{\psi})\right]dv\nonumber\\
I_6&=&
\frac{1}{2}\int_\W\left\{(\psi_2\bar{\psi}_t+\bar{\psi}_2\psi_t)(\hat
\bA_1+\hat \bA_2+2\bA_{\cH})\cdot \hat \bA   \right.\\
&& \left.\ \qquad+
(\psi_2\bar{\psi}_t+\bar{\psi}_2\psi_t)(\bar\psi_1{\psi}+{\psi}_2\bar\psi)-(\psi \bar\psi_{1t}+\bar \psi
\psi_{1t})\hat{u}\right\}dv
\nonumber\\
I_7&=&\int_\W\left[|\psi_1|^2\hat \bA + (\psi\bar{\psi}_1+\psi_2\bar\psi)(\hat
\bA_2+\bA_{\cH}) \right]\cdot(\hat\bA_t-\nabla
\hat u)dv \nonumber\\
I_8&=&\int_\W\left[\frac{i}{2\kappa}({\psi}\nabla\bar\psi_1-\bar{\psi}\nabla\psi_1+{\psi}_2\nabla\bar\psi-\bar{\psi}_2\nabla\psi)\cdot(\nabla \hat u-\hat \bA_t)+\nabla \hat u \cdot
\hat\bA_t\right]dv.
\end{eqnarray*}
By recalling that the solution of \eqref{psi}-\eqref{u2}
satisfies the a-priori estimate \eqref{apriori_z1},  the
previous integrals can be estimated by means of the H\"{o}lder's and Young's
inequalities the Sobolev embedding theorem  as
\begin{eqnarray} \label{stimaI}
\sum_{h=1}^8 I_h &\leq&
\vf_1\|\psi\|_{H^1}^2+\vf_2\|\hat\bA\|^2_{\cH^1}+C\| \hat u\|^2 \nonumber\\
&+&\frac{1}{2}(\gamma\|\psi_t\|^2+k_0\|\nabla \hat u\|^2+\|\hat
\bA_t\|^2)\,,
\end{eqnarray}
where
\begin{eqnarray*}
\varphi_1&=&C_R(1+\|\psi_1\|_{H^2}^2+\|\psi_2\|_{H^2}^2+\|\hat \bA_1\|^2_{\cH^2}+\|\hat\bA_2\|_{\cH^2}^2+\| \hat u_1\|_{H^1_0}^2
+\|\psi_{1t}\|^2)\\
\varphi_2&=&C_R(1+\|\psi_1\|_{H^2}^2+\|\psi_2\|_{H^2}^2).
\end{eqnarray*}

Substitution into (\ref{d}) yields the inequality
\begin{eqnarray*}
&&\frac{1}{2}\frac{d}{dt}\left[\gamma\|\psi\|^2+\frac{1}{\kappa^2}\|\nabla\psi\|^2+
\|\div\hat\bA\|^2+\mu\|\rot\hat\bA\|^2 +c_0\|\hat{u}\|^2\right]
\\
&&\leq
\varphi_1\|\psi\|_{H^1}^2+\varphi_2\|\hat\bA\|_{\cH^1}^2+C\|\hat
u\|^2 .
\end{eqnarray*}
In view of  \eqref{apriori_z2}, \eqref{apriori_z3}, Gronwall's inequality leads to
\begin{eqnarray} \label{Grm1}
&&\|z_1(t) - z_2(t)\|_{\cZ^1}^2\leq C_R e^{C_R t}\,  \| z_{01} -z_{02}\|^2_{\cZ^1} .
\end{eqnarray}

Now we prove inequality (\ref{int_Z2}). We substitute \eqref{stimaI} into \eqref{d} and
integrate over $t$, thus obtaining
\bea
&&\int_0^t \left(\gamma\|\psi_t\|^2 + \|\hat\bA_t\|^2 + k_0 \|\nabla\hat u\|^2 \right)ds \\
&&\leq\int_0^t \left(\varphi_1\|\psi\|_{H^1}^2+\varphi_2\|\hat\bA\|_{\cH^1}^2+C\|\hat u\|^2  \right)ds
+ C\|z_{01} - z_{02}\|_{\cZ^1}^2.
\eea
A-priori estimates (\ref{apriori_z1}), (\ref{apriori_z2}) and inequality \eqref{Grm1} provide
$$
\int_0^t \left(\gamma\|\psi_t\|^2 + \|\hat\bA_t\|^2 + k_0 \|\nabla\hat u\|^2 \right)ds \leq
C(t)\|z_{01} - z_{02}\|_{\cZ^1}^2.
$$
From \eqref{psid} we obtain the estimate of $\|\triangle \psi\|$ by means of H\"older's inequality, \eqref{apriori_z1}-\eqref{apriori_z3} and \eqref{Grm1}, namely
$$
\int_0^t \|\triangle\psi\|^2 ds \leq  C(t)\|z_{01} - z_{02}\|_{\cZ^1}^2.
$$
Likewise, multiplying \eqref{Ad} by $\rot\rot\hat\bA$ and integrating over $\Omega$, we deduce
$$
\|\rot\rot\hat\bA \|^2 \leq C\|\rot\rot\hat\bA\|\, \|\hat\bA_t\| + C(t) \|\rot\rot\hat\bA\| \, \|z_{01} - z_{02}\|_{\cZ^1}^2,
$$
which implies
$$
\int_0^t \|\rot\rot\hat\bA \|^2 ds \leq  C(t)\|z_{01} - z_{02}\|_{\cZ^1}^2.
$$
Finally, by comparison with \eqref{Ad} we reach the conclusion.
\qed

Theorems \ref{exist} and \ref{cont_dep} ensure that there exists a unique solution of problem (\ref{psi})-(\ref{ic}) depending continuously on the initial data. In other words, (\ref{psi})-(\ref{ic}) generate a strongly continuous semigroup
$S(t)$ on the phase space $\cZ^1(\W)$ (see {\it e.g.} \cite{Tem}).

\section{The global attractor}
This section is devoted to prove existence of the global attractor for the semigroup $S(t)$.
For reader's convenience, we recall its definition.
\begin{Def}
The global attractor $\cA \subset \cZ^1(\W)$ is the unique compact set enjoying the following properties:
\begin{itemize}
    \item[(i)] $S(t)\cA = \cA, \qquad \forall t\geq 0$;
    \item[(ii)] $\displaystyle\lim_{t\to \infty} {\rm dist}_{\cZ^1}(S(t)\cB,\cA) =0$ for every bounded set $\cB\subset\cZ^1(\W)$,
    where {\rm dist}$_{\cZ^1}$ denotes the usual Hausdorff semidistance in $\cZ^1(\W)$.
\end{itemize}
\end{Def}

Usually, existence of the global attractor is established by showing that the semigroup admits a bounded absorbing set and that the operators $S(t)$ are uniformly compact for large values of $t$ (\cite[Theor. 1.1]{Tem}). However, we are unable to obtain directly the estimates that guarantee the dissipativity of the semigroup. This prevent us from proving existence of an absorbing set. Thus we deduce that $S(t)$ possesses a global attractor by means of a Lyapunov functional which leads to existence of a bounded absorbing set as a consequence.

We denote by $\cS$ the set of stationary solutions of problem (\ref{psi})-(\ref{bc2}). In other words, every steady solution  satisfies the equations
\begin{eqnarray}
\label{psi_staz}
0&=&\frac{1}{\kappa ^{2}}\triangle \psi-\frac{2i}{\k}(\hat\bA+\bA_\cH)\cdot\nabla\psi-\psi|\hat\bA+\bA_\cH|^2+i\b(\div\hat\bA)\psi
\nonumber\\
&&-\psi(|\psi|^{2}-1+\hat u + u_\cH)
 \\
 \label{A_staz}
{\bf 0}&=&\nabla(\div\hat\bA)-\mu\rot\rot\hat\bA- |\psi|^2(\hat\bA+\bA_\cH)+\frac{i}{2\k}(\psi\nabla\bar{\psi}-\bar{\psi}\nabla\psi)
\nonumber\\
&&-\nabla \hat u- \rot\bG\\
\label{u2_staz}
0&=&k_0\triangle \hat u +
\div\left[-|\psi|^2(\hat\bA+\bA_\cH)+\frac{i}{2\k}(\psi\nabla\bar\psi-\bar\psi\nabla\psi)\right]
\end{eqnarray}

\begin{Def}
A continuous function $\cL:\cZ^1(\W)\to \R$ is said a Lyapunov functional if
\begin{itemize}
    \item[(i)] $t\to\cL(S(t)z) \text{ is non-increasing for any } z \in \cZ^1(\W)$;
    \item[(ii)] $\cL(z)\to \infty \Leftrightarrow \|z\|_{\cZ^1} \to \infty$;
    \item[(iii)] $\cL(S(t)z) = \cL(z), \ \forall t > 0 \Rightarrow z\in \cS.$
\end{itemize}
\end{Def}

In order to prove the existence of the global attractor, we will exploit the following result  (see {\it e.g.} \cite{BV,Hale}).
\begin{theor}\label{glob}
Let the semigroup $S(t)$, $t>0$ satisfy the following conditions:
\begin{itemize}
    \item[(a)] $S(t)$ admits a continuous Lyapunov functional $\cL$;
    \item[(b)] the set $\cS$ of the stationary solutions is bounded in $\cZ^1(\W)$;
    \item[(c)] for any bounded set $\cB \subset \cZ^1(\W)$, there exists a compact set $\cK_{\cB} \subset \cZ^1(\W)$ such that $S(t)\cB \subset \cK_{\cB}$, $t>0$.
\end{itemize}
Then, $S(t)$ possesses a connected global attractor $\cA$ which coincides with the unstable manifold of $\cS$, namely
\bea
\cA &=& \{z \in \cZ^1(\W) :  z\ belongs\ to\ a\ complete\ trajectory\ S(t)z,\ t\in \R,\\
&&\ \lim_{t\to-\infty} {\rm dist}_{\cZ^1}(S(t)z,\cS)=0\}.
\eea
\end{theor}
The next subsections will be devoted to the proof of conditions ($a$), ($b$), ($c$).

\subsection{Lyapunov functional}
\begin{prop}\label{Lyfu}
The function
\begin{eqnarray*}
    \cL(\psi,\hat\bA,\hat u) &=& \frac12\int_\W \Bigg\{\left|\frac{i}{\k}\nabla\psi + \psi\hat\bA +\psi \bA_\cH\right|^2 + \frac{1}{2}(|\psi|^2-1)^2 +  |\psi|^2u_\cH 
    \\
    &&+ \mu|\rot\hat\bA|^2+ \eta(\div\hat\bA)^2 +2\rot\bG \cdot \hat\bA + c_0\hat u^2 \Bigg\} dv,
\end{eqnarray*}
where $\eta=2k_0/(k_0+1)$, is a Lyapunov functional.
\end{prop}

\noindent
{\bf Proof.}
The non-increasing character of $\cL$ has been proved in proposition \ref{a-priori}. Moreover, the inequalities
\bea
&&c_1 \cF_1 - c_2 \leq \cL \leq c_3\cF_1 + c_2
\\
&&\|z\|_{\cZ^1}^2\leq C\left[1+\cF_1(z)+\cF_1^2(z)+\cF_1^3(z)\right]\\
\eea
hold. With similar arguments one can show that
$$
\cF_1(z(t))\leq C(1+\|z(t)\|^2_{\cZ^1}+\|z(t)\|^4_{\cZ^1}).
$$
Hence, we deduce that
$$
\cF_1(z)\to \infty \quad \Leftrightarrow \quad \cL(z)\to \infty \quad \Leftrightarrow\quad  \|z\|_{\cZ^1} \to \infty.
$$

Finally, we show ({\it iii}). We suppose that $\cL(S(t)z)=\cL(z)$
for every $t>0$. Then, from (\ref{Lq}) and the positive definitess
of $q$ we deduce that
\begin{eqnarray}
\label{f1}
\psi_t-i\k\psi\div\hat\bA & = & 0 \\
\label{grad_div_A}
\nabla(\div\hat\bA ) & = & {\bf 0}\\
\label{u1}
\nabla\hat u & = & {\bf 0}\\
\hat\bA_t &=& {\bf 0}
\end{eqnarray}
In particular, (\ref{grad_div_A}) guarantees that there exists a constant $c$ such that $\div\hat\bA = c$.
Since $\div\hat\bA\in H^1_{0m}(\W)$, we have
$$
\div\hat\bA=0,
$$
which, in view of (\ref{f1}) implies
\begin{equation}\label{psit}
\psi_t = 0.
\end{equation}
Finally, by substituting the previous relations into (\ref{A}), (\ref{u2}), we obtain
$$
\hat u_t = 0.
$$
Thus, $z_t=0$, namely $z\in \cS$.
\qed


\subsection{Stationary solutions}
\begin{prop}
The set of stationary solutions is bounded in $\cZ^1(\W)$, namely there exists $R>0$ such that
$$
\|z\|_{\cZ^1}\leq R,
$$
for every $z\in\cS$.
\end{prop}

\noindent
{\bf Proof.}
Let $z \in \cS$. Then, $\displaystyle\frac{d\cL}{dt} =0$ and hence
\begin{eqnarray*}
\nabla\hat u  =  {\bf 0}, \qquad\qquad
\nabla(\div\hat\bA)  =  {\bf 0}.
\end{eqnarray*}
In particular, the boundary conditions (\ref{bc1}) and (\ref{bc2}) lead to $\hat u =0$ and $\div\hat\bA=0$. By substituting into (\ref{psi_staz})-(\ref{A_staz}) we obtain
\begin{eqnarray}
\label{psi_staz1}
&&\frac{1}{\kappa ^{2}}\triangle \psi-\frac{2i}{\k}(\hat\bA+\bA_\cH)\cdot\nabla\psi-\psi|\hat\bA+\bA_\cH|^2
-\psi(|\psi|^{2}-1 + u_\cH) =0\qquad
 \\
 \label{A_staz1}
&&\mu\rot\rot\hat\bA+ |\psi|^2(\hat\bA+\bA_\cH)-\frac{i}{2\k}(\psi\nabla\bar{\psi}-\bar{\psi}\nabla\psi)
+ \rot\bG={\bf 0}\qquad
\end{eqnarray}
By multiplying (\ref{psi_staz1}) by $1/2\bar\psi$, its conjugate by $1/2\psi$ and integrating over $\W$ we obtain
$$
\left\|\frac{i}{\k}\nabla \psi+(\hat\bA+\bA_\cH)\psi\right\|^2 +
\|\psi\|_4^4 = \|\psi\|^2- \int_\W |\psi|^2u_\cH dv \ ;
$$
H\"older's inequality yields
\begin{equation}\label{hold}
    \|\psi\|^2 \leq \e \|\psi\|^4_4 + C,
\end{equation}
where $\e>0$ is a suitable (small) constant.
Therefore, we have
\begin{eqnarray}
\label{psi4}
\|\psi\|_4 & \leq & C, \\
\label{nabla}
\left\|\frac{i}{\k} \nabla\psi + (\hat\bA  + \bA_\cH) \psi \right\| & \leq & C.
\end{eqnarray}
We multiply (\ref{A_staz1}) by $\hat\bA$ and we integrate over $\W$, thus obtaining
\begin{eqnarray*}
\mu\|\rot\hat\bA\|^2 &=&  -\int_\W\left[ |\psi|^{2}(\hat\bA+\bA_\cH) -\frac{i}{2\k}(\psi\nabla\bar{\psi}-\bar{\psi}\nabla\psi)+ \rot\bG
\right]\cdot\hat\bA dv
\\
&=&-\int_\W\left\{ \frac{1}{2}\left[\frac{i}{\k} \nabla\psi + (\hat\bA  + \bA_\cH) \psi \right]\bar\psi\right.\\
&&\left.+
\frac{1}{2}\left[-\frac{i}{\k} \nabla\bar\psi + (\hat\bA  + \bA_\cH) \bar\psi \right]\psi+ \rot\bG
\right\}\cdot\hat\bA dv\\
& \leq & \left\|\frac{i}{\k}\nabla \psi+(\hat\bA+\bA_\cH)\psi\right\| \|\psi\hat\bA\| +  \|\rot\bG\|\, \|\hat\bA\|
\end{eqnarray*}
Hence thanks to (\ref{psi4}) and (\ref{nabla}) we deduce
$$
\mu\|\rot\hat\bA\|^2\leq C \|\hat\bA\|_4+\|\rot\bG\| \|\hat\bA\|\leq C\|\hat\bA\|_{\cH^1}=C\|\rot\hat\bA\|,
$$
where last identity holds since $\div\hat\bA =0$. Thus,
\begin{equation}\label{A_lim}
    \|\hat\bA\|_{\cH^1} \leq C.
\end{equation}
Finally, in view of the inequalities (\ref{hold})-(\ref{A_lim}) we obtain
\begin{eqnarray*}
\|\psi\| & \leq & C \\
\frac{1}{\k} \|\nabla\psi\| & \leq & \left\|\frac{i}{\k} \nabla\psi + (\hat\bA  + \bA_\cH) \psi \right\|
+ \left\| \hat\bA + \bA_\cH \right\|_4 \|\psi \|_4 \leq C,
\end{eqnarray*}
namely
$$
\|\psi\|_{H^1}\leq C.
$$
This concludes the proof.
\qed

\subsection{Existence of the global attractor}
Existence of the global attractor for the semigroup $S(t)$ is established once we prove condition ($c$) of Theorem \ref{glob}.

\begin{prop}\label{prop_lim}
Let $S(t)z$, $t>0$, be a solution to problem (\ref{psi})-(\ref{ic}) with initial datum $z \in \cZ^1(\W)$ such that
$\|z\|_{\cZ^1} \leq R$. Then, $S(t)z$, $t>0$, belong to a compact set $\cK \subset \cZ^1(\W)$.
\end{prop}

\noindent
{\bf Proof.}
In view of the compact embedding $\cZ^2(\W) \hookrightarrow \cZ^1(\W)$, our goal consists in proving the existence of a positive constant $C_R$ depending on $R$ and $\bA_{\cH}, u_\cH, \bG$ such that
\begin{equation}\label{comp}
    \| z(t) \|_{\cZ^2} \leq C_R.
\end{equation}

Let us multiply (\ref{psi}) by $1/2\, \triangle \bar\psi_t$ and its conjugate by $1/2\, \triangle\psi_t$. Adding the resulting equations and integrating over $\W$, thanks to the boundary condition (\ref{bc2})$_1$, we obtain
\bea
\frac{1}{2\k^2}\frac{d}{dt}\|\triangle\psi\|^2 + \g \|\nabla\psi_t\|^2 = \sum_{h=1}^4 J_h,
\eea
where
\bea
J_1 &=& \frac{i}{\k} \int_\W (\hat\bA + \bA_\cH)\cdot (\nabla\psi \triangle\bar\psi_t - \nabla\bar\psi \triangle\psi_t)\,dv
\\
J_2 &=& \frac{1}{2}\int_\W |\hat\bA + \bA_\cH|^2 (\bar\psi \triangle\psi_t + \psi \triangle\bar\psi_t)\,dv
\\
J_3 &=& \frac{i\b}{2} \int_\W \div\hat\bA\, (\bar\psi \triangle\psi_t - \psi \triangle\bar\psi_t)\,dv
\\
J_4 &=& \frac{1}{2}\int_\W (|\psi|^2 -1 +\hat u +u_\cH) (\bar\psi \triangle\psi_t + \psi \triangle\bar\psi_t)\,dv.
\eea
An integration by parts leads to
$$
|J_1| \leq C \int_\W \left[|\nabla(\hat\bA + \bA_\cH)| |\nabla\psi| + |\nabla\nabla\psi| |\hat\bA + \bA_\cH| \right]
|\nabla\psi_t| \, dv.
$$
H\"older's and Young's inequalities  and (\ref{Sob2}) yield
\bea
|J_1| &\leq & C\|\nabla\psi_t\| \left[ \|\hat\bA + \bA_\cH\|_{\cH^2} \|\psi\|_{H^2} + \|\hat\bA + \bA_\cH\|_\infty \|\psi\|_{H^2} \right]
\\
&\leq & \e \|\nabla\psi_t\|^2 + \frac{C}{4\e} \|\hat\bA + \bA_\cH\|_{\cH^2}^2 \|\psi\|^2_{H^2},
\eea
for any $\e>0$.

Now we consider $J_2$. We obtain
\bea
|J_2| &\leq &  \int_\W|\nabla\psi_t| \left[2 |\nabla(\hat\bA + \bA_\cH)| |\hat\bA + \bA_\cH| |\psi|
 + |\hat\bA + \bA_\cH|^2  |\nabla\psi| \right] \, dv
\\
&\leq & C \|\nabla\psi_t\| \left[ \|\hat\bA + \bA_\cH\|_{\infty}\|\psi\|_\infty \|\hat\bA + \bA_\cH\|_{\cH^1}
+ \|\hat\bA + \bA_\cH\|_{\infty}^2 \|\nabla\psi\| \right].
\eea
The assumption $\|z\|_{\cZ^1}\leq R$ together with (\ref{apriori_z1}) give
\bea
|J_2| &\leq & C_R\|\nabla\psi_t\| \left[ \|\hat\bA + \bA_\cH\|_{\infty}\|\psi\|_\infty
+ \|\hat\bA + \bA_\cH\|_{\infty}^2 \right]
\\
&\leq & \e \|\nabla\psi_t\|^2 + \frac{C_R}{4\e} \|\hat\bA + \bA_\cH\|_{\cH^2}^2 (\|\psi\|^2_{H^2}+\|\hat\bA + \bA_\cH\|_{\cH^2}^2).
\eea

Similarly, we have
\bea
|J_3| &\leq& \e \|\nabla\psi_t\|^2 + \frac{C}{4\e} \|\hat\bA + \bA_\cH\|_{\cH^2}^2 \|\psi\|^2_{H^2}\\
|J_4| &\leq &  \e \|\nabla\psi_t\|^2 + \frac{C_R}{4\e}  (\|\psi\|^4_{H^2}+\|\psi\|_{H^2}^2\|\hat u\|_{H^1_0}^2+1).
\eea
Therefore,
\begin{eqnarray}\nonumber
&&  \frac{1}{2\k^2}\frac{d}{dt}\|\triangle\psi\|^2 + \g \|\nabla\psi_t\|^2\leq 4 \e \|\nabla\psi_t\|^2 
    \\\label{J}
&&  + C_R \left[ \|\hat\bA + \bA_\cH\|_{\cH^2}^2\|\psi\|^2_{H^2}  +\|\hat\bA + \bA_\cH\|_{\cH^2}^4 + \|\psi\|^4_{H^2}+ \|\psi\|_{H^2}^2\|\hat u\|_{H^1_0}^2+ 1\right].\nonumber\\
\end{eqnarray}

Let us multiply (\ref{A}) by $\rot\rot\hat\bA_t$. Keeping (\ref{bc1})$_2$ into account, an integration over $\W$ provides
$$
\frac{\mu}{2} \frac{d}{dt} \|\rot\rot\hat\bA\|^2 + \|\rot\hat\bA_t\|^2 = \sum_{h=1}^3 L_h,
$$
where
\bea
L_1 &=& -\int_\W \rot[|\psi|^2 (\hat\bA +\bA_\cH)]\cdot \rot\hat\bA_t \, dv
\\
L_2 &=& \frac{i}{2\k}\int_\W \rot(\psi\nabla\bar\psi -\bar\psi\nabla\psi) \cdot \rot\hat\bA_t \, dv
\\
L_3 &=& -\int_\W \rot\rot\bG \cdot \rot\hat\bA_t \, dv.
\eea
Thus, we obtain
\bea
|L_1| & \leq& \e\|\rot\hat\bA_t\|^2 +\frac{C_R}{4\e} \left[ \|\psi\|^2_{H^2} + \|\hat\bA +\bA_\cH\|^2_{\cH^2}\right]
\\
|L_2| &\leq & \e\|\rot\hat\bA_t\|^2 + \frac{C}{4\e} \|\psi\|^4_{H^2}
\\
|L_3| &\leq&  \e\|\rot\hat\bA_t\|^2 + \frac{1}{4\e}\|\rot\rot\bG\|^2.
\eea
Therefore,
\begin{equation}\label{L}
    \frac{\mu}{2} \frac{d}{dt} \|\rot\rot\hat\bA\|^2 + \|\rot\hat\bA_t\|^2
    \leq 3\e \|\rot\hat\bA_t\|^2 + C_R \left(\|\psi\|^4_{H^2}+ \|\hat\bA +\bA_\cH\|^2_{\cH^2} +1 \right).
\end{equation}

Let us multiply (\ref{A}) by $\nabla(\div \hat\bA_t) - \nabla \hat u_t$. An integration by parts and boundary conditions (\ref{bc1})-(\ref{bc2}) lead to
$$
\frac{1}{2}\frac{d}{dt}\|\nabla(\div\hat\bA) - \nabla\hat u -\rot\bG\|^2 + \|\div\hat\bA_t \|^2 = \sum_{h=1}^4 M_h
$$
with
\bea
M_1 &=&  \int_\W \hat u_t\div\hat\bA_t \,dv
\\
M_2 &=& -\int_\W \left[\nabla(|\psi|^2)\cdot(\hat\bA + \bA_\cH) +  |\psi|^2\div\hat\bA\right]\, \div\hat\bA_t\,dv
\\
M_3 &=& \frac{i}{2\k} \int_\W (\psi\triangle\bar\psi - \bar\psi \triangle\psi)\,\div\hat\bA_t\,dv
\\
M_4 &=& \int_\W \left[ \nabla(|\psi|^2)\cdot (\hat\bA+\bA_{\cH})+|\psi|^2\div\hat\bA
-\frac{i}{2\k}(\psi\triangle\bar{\psi}-\bar{\psi}\triangle\psi)
\right]  \hat u_t \,dv.
\eea
We obtain
\bea
|M_1| &\leq&  \e\|\div\hat\bA_t\|^2 + \frac{1}{4\e}\|\hat u_t\|^2
\\
|M_2| &\leq& \e\|\div\hat\bA_t\|^2 + \frac{C_R}{4\e}
\left[\|\hat\bA + \bA_\cH\|^2_{\cH^2} +  \|\psi\|^2_{H^2}\right]
\\
|M_3| &\leq& \e\|\div\hat\bA_t\|^2 + \frac{C}{4\e}\|\psi\|^4_{H^2}
\\
|M_4| &\leq& \e\|\hat u_t\|^2 +  \frac{C_R}{4\e} \left[\|\hat\bA +
\bA_\cH\|^2_{\cH^2} +  \|\psi\|^4_{H^2}\right]. \eea Therefore, we
have
\begin{eqnarray}\nonumber
    &&
    \frac{1}{2}\frac{d}{dt}\|\nabla(\div\hat\bA) - \nabla\hat u -\rot\bG\|^2 + \|\div\hat\bA_t \|^2\leq
    3\e\|\div\hat\bA_t\|^2
    \\\label{M}
    &&
     + \left(\e+\frac{1}{4\e} \right)\|\hat u_t \|^2 + {C_R}
    \left[1+\|\hat\bA+\bA_\cH\|^2_{\cH^2}+\|\psi\|^4_{H^2}  \right].
\end{eqnarray}

Let us multiply (\ref{u2}) by $\hat u_t$ and integrate over $\W$.
\bea
\frac{k_0}{2}\frac{d}{dt}\|\nabla\hat u\|^2 + c_0\|\hat u_t\|^2 = N_1 + N_2
\eea
with
\bea
N_1 &=&  \frac12\int_\W (\psi_t\bar\psi + \bar\psi_t\psi) u_t\,dv
\\
N_2 &=& \int_\W \div\left[-|\psi|^2(\hat\bA+\bA_\cH)+\frac{i}{2\k}(\psi\nabla\bar\psi-\bar\psi\nabla\psi)\right]
\hat u_t\,dv
\eea
H\"older's, Young's inequalities and (\ref{Sob2}) yield
$$
|N_1| \leq \e \|\hat u_t\|^2 + \frac{C}{4\e}\|\psi\|^2_{H^2}\|\psi_t\|^2.
$$
Since $N_2 =- M_4$, we deduce that
\begin{equation}\label{N}
    \frac{k_0}{2}\frac{d}{dt}\|\nabla\hat u\|^2 + c_0\|\hat u_t\|^2
    \leq
    2\e \|\hat u_t\|^2 + C_R \left[\|\psi\|^2_{H^2}\|\psi_t\|^2 +
    \|\psi\|^4_{H^2} +\|\hat\bA+\bA_\cH\|^2_{\cH^2}\right].
\end{equation}
We multiply (\ref{N}) by $1/(2\e c_0)$.
By adding the resulting inequality with (\ref{J})-(\ref{M}), we obtain
\begin{eqnarray}\nonumber
&&
    \frac12 \frac{d}{dt} \left[\frac{1}{\k^2}\|\triangle\psi \|^2 +
    \mu \|\rot\rot\hat\bA\|^2 +
    \|\nabla(\div\hat\bA) - \nabla\hat u -\rot\bG\|^2 \right. \\ \nonumber
    &&\left.
    + \frac{k_0}{2\e c_0} \|\nabla\hat u\|^2  \right]
    + (\g-4\e) \|\nabla\psi_t\|^2 + (1-3\e)\|\rot\hat\bA_t\|^2 
    \\ \label{conti}
    &&+ (1-3\e)\|\div\hat\bA_t \|^2+ \left(\frac{1}{4\e}-\e -\frac{1}{c_0} \right)\|\hat u_t\|^2\nonumber\\
    &&\leq
C_R \left[ 1+\|\hat\bA + \bA_\cH\|_{\cH^2}^2\|\psi\|^2_{H^2}  +\|\hat\bA + \bA_\cH\|_{\cH^2}^4 + \|\psi\|^4_{H^2}+
        \|\psi\|_{H^2}^2\|\hat u\|_{H^1_0}^2\right.\nonumber\\
    &&\left.+\|\psi\|^2_{H^2}\|\psi_t\|^2 \right]
\end{eqnarray}
We choose
$$
\e = \frac12 \text{min} \left(\frac{\g}{4}, \frac13, \frac{\sqrt{1+c_0^2}-1}{2c_0} \right)
$$
and we let
\bea
\cF_2 & = & \frac{1}{\k^2}\|\triangle\psi \|^2 + \mu \|\rot\rot\hat\bA\|^2 +
    \|\nabla(\div\hat\bA) - \nabla\hat u -\rot\bG\|^2 \\
    && + \frac{k_0}{2\e c_0} \|\nabla\hat u\|^2
\\
\xi(t) &= & C_R \left[1+\|\psi\|^2_{H^2} + \|\psi_t\|^2 + \|\hat\bA\|_{\cH^2}^2 \right]
\eea
Therefore, we have
\begin{equation}\label{F'}
    \frac{d}{dt}\cF_2 \leq \xi(t)\cF_2 + \xi(t).
\end{equation}
A-priori estimates (\ref{apriori_z1})-(\ref{apriori_z3}) and Gronwall's uniform lemma (\cite{Tem}) guarantee that $\cF_2$ is bounded.

Thus, $\|z(t)\|_{\cZ^2}<C_R$.
\qed

\begin{rem}\label{remark}
{\rm By comparison with (\ref{psi}), on account of the H\"older's inequality and the Sobolev embedding theorem, from (\ref{comp}) we prove
\bea
\|\psi_t\|&\leq& C(\|\triangle\psi\|+\|\hat\bA+\bA_\cH\|_{\cH^1}\|\psi\|_{H^2}+\|\hat\bA+\bA_\cH\|_{\cH^1}^2\|\psi\|_{H^1}+\|\psi\|_{H^1}^3\\
&&+\|\psi\|_{H^2}+\|\hat u +u_\cH\|\|\psi\|_{H^2})\\
&\leq& C_R.
\eea
}
\end{rem}

\medskip

Propositions \ref{Lyfu}-\ref{prop_lim} allow to apply Theorem (\ref{glob}) and to prove existence of the global attractor.
As a consequence (\cite{CP}), $S(t)$ possesses a bounded absorbing set $\mathcal B_1 \subset \cZ^1(\W)$ of radius
$$R_1=1+\sup\{\|z\|_{\cZ^1},\ \cL(z)\leq K\},$$
where $K=1+\displaystyle\sup_{z\in\cS}\cL(z)$.

\begin{cor}\label{corass}
The semigroup $S(t)$ possesses a bounded absorbing set $\cB_2\in\cZ^2(\W)$ of radius $R_2$.
\end{cor}

\noindent
{\bf Proof.}
Let $z\in\cZ^2(\W)$ with $\|z\|_{\cZ^2}\leq R$. Then there exists $t_1=t_1(R)>0$ such that
$$
S(t)z\in \cB_1,\qquad t\geq t_1,
$$
so that
$$
\|S(t)z\|_{\cZ^1}\leq R_1,\qquad t\geq t_1.
$$
Inequality (\ref{comp}) implies
$$
\|S(t)z\|_{\cZ^2}=\|S(t-t_1)S(t_1)z\|_{\cZ^2}\leq C_{R_1},\qquad t\geq t_1.
$$
If $t<t_1$, the same inequality (\ref{comp}) leads to
$$
\|S(t)z\|_{\cZ^2}\leq C_{R}e^{t_1-t}.
$$
Therefore, we obtain
$$
\|S(t)z\|_{\cZ^2}\leq  C_{R_1} +C_{R}e^{t_1-t},\qquad t> 0.
$$
By choosing $R_2=2C_{R_1}$ and $t_2=\max\left\{t_1 - \ln(C_{R_1}/C_R),0\right\}$, we prove
$$
\|S(t)z\|_{\cZ^2}\leq  R_2,\qquad t> t_2.
$$
\qed

\section{Exponential attractors}
In this section, we prove the existence of a regular exponential attractor $\cE$ for the semigroup $S(t)$, namely, a compact set of finite fractal dimension that exponentially attracts every bounded set in $\cZ^2(\W)$. Since the global attractor $\cA$ is the minimal compact attracting set, we have $\cA \subset \cE$. Accordingly, $\cA$ has finite fractal dimension.

We first recall the definition of the exponential attractor
\begin{Def}
A compact subset $\cE\subset\cZ^2(\W)$ of finite fractal dimension is an exponential attractor for the semigroup $S(t)$ if
\begin{itemize}
\item[(i)] $\cE$ is positively invariant, {\it i.e.} $S(t)\cE\subset\cE$ for all $t\geq 0$;
\item[(ii)] there exist  $\w>0$ and a positive increasing function $J$ such that
\begin{equation}\label{exp_rate}
{\rm dist}_{\cZ^1}(S(t)\cB,\cE)\leq J(R) e^{-\w t}
\end{equation}
for any bounded $\cB\subset\cZ^1(\W)$ with $R=\sup\{\|z\|_{\cZ^1(\W)}, z\in\cB\}$.
\end{itemize}
\end{Def}
The existence of an exponential attractor for the semigroup $S(t)$ is based on the following abstract result proved in \cite{GGMP}.
\begin{lem}\label{lemma1}
Let $\cK$ a bounded subset of $\cZ^2(\W)$, such that $S(t)\cK \subset \cK$ for each $t>t^*$. Suppose that
\begin{itemize}
    \item[(i)] the map
    \bea
 \Phi: [t^*,2t^*] \times \cK &\rightarrow& \cK
    \\
    (t,z) &\mapsto& S(t)z
    \eea
    is $1/2$-H\"{o}lder continuous in time and Lipschitz continuous in the initial data, when $\cK$ is endowed with the     $\cZ^1(\W)$-topology;
    \item[(ii)] there exist $\l\in (0,1/2)$ and $\L>0$ such that
    $$
    S(t^*) = L + K,
    $$
    where
    \bea
    \|L(z_1)-L(z_2)\|_{\cZ^1} &\leq& \l \|z_1 -z_2\|_{\cZ^1},
    \\
    \|K(z_1)-K(z_2)\|_{\cZ^2} &\leq& \L \|z_1 -z_2\|_{\cZ^1},    \qquad z_1, z_2 \in \cK.
    \eea
    \end{itemize}
Then, there exists a set $\cE \subset \cK$, closed and of finite fractal dimension in $\cZ^1(\W)$, positively invariant for $S(t)$, such that
\begin{equation}\label{exp}
\text{dist}_{\cZ^1} (S(t)\cK, \cE) \leq J_0 e^{-\w t},
\end{equation}
for some $\w>0$, $J_0\geq 0$.
\end{lem}

In order to prove that $\cE$ is an exponential attractor for the semigroup $S(t)$, we have to show that the condition (\ref{exp}) holds replacing $\cK$ with an arbitrary bounded set $\cB\subset\cZ^1(\W)$.
To this aim, we prove in the following lemma \ref{lemma2} that the absorbing set $\cB_2$ exponentially attracts every bounded set $\cB\subset\cZ^1(\W)$. Accordingly, owing to the transitivity property of exponential attraction, we prove the existence of an exponential attractor for $S(t)$.

\begin{lem}\label{lemma2}
The absorbing set $\cB_2 \subset \cZ^2(\W)$ satisfies the following conditions
\begin{itemize}
    \item[(i)] there exists an increasing function $M$ such that for every bounded set $\cB \subset \cZ^1(\W)$ we have
    \begin{equation}\label{expB2}
    \text{dist}_{\cZ^1} (S(t)\cB, \cB_2) \leq M(R) e^{-\nu t}
    \end{equation}
    where $R= \displaystyle \sup_{z\in\cB} \|z \|_{\cZ^1}$ and $\nu$ is a positive constant independent of $R$;
    \\
    \item[(ii)] there exists $ t_2>0$ such that
    $$
    S(t)\cB_2 \subset \cB_2, \qquad\qquad \forall t\geq t_2.
    $$
\end{itemize}
\end{lem}
\noindent
{\bf Proof.}
Let us split the solution $S(t)z=z(t)$ as the sum
$$
z(t)=z^l(t)+z^k(t),
$$
where $z^l(t)=(\psi^l(t), \hat\bA^l(t), \hat u^l(t))$  solves the differential problem
\begin{eqnarray}\label{psi_l}
&&\g\psi_t^l-\frac{1}{\k^2}\triangle\psi^l+\psi^l=0\\
\label{A_l}
&&\hat\bA_t^l-\nabla(\div\hat\bA^l)+\mu\rot\rot\hat\bA^l + \nabla\hat u^l=0\\
\label{u_l}
&&c_0\hat u_t^l-k_0\triangle\hat u^l=0\\
\label{ic_l}
&&\nabla\psi^l\cdot\bn=0,\ \ \hat\bA^l\cdot\bn=0,\ \ (\rot\hat\bA^l)\times\bn={\bf 0},\ \ \hat u^l=0,\ \ {\rm on\ }\partial\W\\
\label{bc_l}
&& \psi^l(0)=\psi_0,\qquad \hat\bA^l(0)=\hat\bA_0,\qquad \hat u^l(0)=\hat u_0
\end{eqnarray}
with $z^l(0)=z(0)=(\psi_0, \hat\bA_0, \hat u_0)$
$$
\|z^l(0)\|_{\cZ^1} \leq R.
$$
Moreover, $z^k(t)=(\psi^k(t), \hat\bA^k(t), \hat u^k(t))$ is a solution to
\begin{eqnarray}\label{psi_k}
&&\g\psi_t^k-\frac{1}{\k^2}\triangle\psi^k+\psi^k=\U(\psi, \hat\bA, \hat u)\\
\label{A_k}
&&\hat\bA_t^k-\nabla(\div\hat\bA^k)+\mu\rot\rot\hat\bA^k + \nabla\hat u^k =\Theta(\psi, \hat\bA, \hat u)\\
\label{u_k}
&&c_0\hat u_t^k-k_0\triangle\hat u^k=\G(\psi, \hat\bA, \hat u)\\
\label{ic_k}
&&\nabla\psi^k\cdot\bn=0,\ \ \hat\bA^k\cdot\bn=0,\ \ (\rot\hat\bA^k)\times\bn={\bf 0},\ \ \hat u^k=0,\  {\rm on\ }\partial\W\ \\
\label{bc_k}
&& \psi^k(0)=0,\qquad \hat\bA^k(0)={\bf 0},\qquad \hat u^k(0)= 0
\end{eqnarray}
where $\U,\Theta,\G$ are defined as
\bea
\U&=&-\frac{2i}{\k}(\hat\bA+\bA_\cH)\cdot\nabla\psi-\psi|\hat\bA+\bA_\cH|^2+i\b\psi\div\hat\bA\\
&&-\psi(|\psi|^2-2+\hat u+u_\cH)\\
\Theta&=&-|\psi|^2(\hat \bA+\bA_\cH)+\frac{i}{2\k}(\psi\nabla\bar\psi-\bar\psi\nabla\psi)-\rot\bG\\\
\G&=&\frac{1}{2}(\psi_t\bar\psi+\bar\psi_t\psi)+\div\left[-|\psi|^2(\hat\bA+\bA_\cH)+\frac{i}{2\k}(\psi\nabla\bar\psi-\bar\psi\nabla\psi)\right]\\
\eea

We prove that
\begin{equation}\label{zc}
\|z^l(t)\|_{\cZ^1}\leq m_1(R) e^{-\nu t}.
\end{equation}
To this aim, let us multiply (\ref{psi_l}) by $\frac{1}{2}(\bar\psi^l_t +\bar\psi^l)$, its conjugate by $\frac{1}{2}(\psi^l_t + \psi^l)$,  (\ref{A_l}) by $\hat\bA_t^l + \hat\bA^l$ and (\ref{u_l}) by $\sigma\hat u^l$, where $\sigma$ is a positive constant large enough. Adding the resulting equations and intergrating over $\Omega$ we obtain
\bea
&&
\frac{1}{2}\frac{d}{dt}\bigg[\frac{1}{\k^2}\|\nabla\psi^l\|^2 + (\g+1)\|\psi^l\|^2 + \|\div\hat\bA^l\|^2 + \mu\|\rot\hat\bA^l\|^2 + \|\hat\bA^l\|^2\\
&& +c_0\sigma \|\hat u^l\|^2 \bigg]
+ \g\|\psi_t^l\|^2 + \frac{1}{\k^2}\|\nabla\psi^l\|^2 + \|\psi^l\|^2 + \|\hat\bA^l_t\|^2 + \|\div\hat\bA^l\|^2\\
&& + \mu\|\rot\hat\bA^l\|^2 + k_0\sigma \|\nabla\hat u^l\|^2
\leq \frac12 (\|\hat\bA_t^l \|^2 + \|\hat\bA^l \|^2_{\cH^1}) + C\|\nabla\hat u^l\|^2.
\eea
Gronwall's inequality implies the existence of a suitable constant $\nu>0$ independent of $R$ and of an increasing function $m_1(R)$ such that (\ref{zc}) holds.

Now let us prove that $z^k(t)$ belongs to a bounded set $\tilde\cB_2\subset\cZ^2(\W)$, namely
\begin{equation}\label{zk}
\|z^k(t)\|_{\cZ^2}\leq m_2(R).
\end{equation}
Let us multiply (\ref{psi_k}) by $-1/2(\triangle \bar\psi_t^k + \triangle \bar\psi^k)$ and its conjugate by
$-1/2( \triangle\psi_t^k + \triangle\psi^k)$. An integration over $\W$ leads to
\bea
&&
\frac12\frac{d}{dt}\left[\frac{1}{\k^2}\|\triangle\psi^k \|^2 + (\gamma + 1)\|\nabla\psi^k\|^2\right]
+\frac{1}{\k^2}\|\triangle\psi^k \|^2 + \gamma\|\nabla\psi^k_t\|^2 + \|\nabla\psi^k \|^2
\\
&&
 \leq \int_\W \left[\left|\nabla\psi_t^k \right| \left|\nabla\Upsilon\right| +
 \left|\triangle\psi^k \right| \left|\Upsilon\right| \right]dv.
\eea
H\"older's and Young's inequalities assure that
\begin{eqnarray}\label{Uk}
    &&\frac{d}{dt}\left[\frac{1}{\k^2}\|\triangle\psi^k \|^2 + (\gamma + 1)\|\nabla\psi^k\|^2\right]
    +\frac{1}{\k^2}\|\triangle\psi^k \|^2 + \gamma\|\nabla\psi^k_t\|^2 + 2\|\nabla\psi^k \|^2\nonumber \\
    &&\leq C \left\|\Upsilon\right\|_{H^1}^2.
\end{eqnarray}
Next, the product in $L^2(\W)$ of (\ref{A_k}) by $\rot\rot\hat\bA_t^k +\rot\rot\hat\bA^k$ yields the inequality
\begin{equation}\label{Tk}
    \frac{d}{dt}\left[\mu \|\rot\rot\hat\bA^k\|^2 + \|\rot\hat\bA^k\|^2 \right]+
    \mu \|\rot\rot\hat\bA^k\|^2 + \|\rot\hat\bA^k_t\|^2  \leq C \|\Theta\|^2_{H^1}.
\end{equation}
Similarly, by multiplying in $L^2(\W)$ the same equation
(\ref{A_k}) by $-\nabla(\div\hat\bA_t^k) + \nabla \hat u_t^k  -
\nabla(\div\hat\bA^k) + \nabla \hat u^k $, we infer
\begin{eqnarray}
    &&\frac{d}{dt}\left[\|-\nabla(\div\hat\bA^k) + \nabla\hat u^k\|^2 + \|\div\hat\bA^k\|^2 \right] \nonumber\\
\label{Tk2}
 && +
    \|-\nabla(\div\hat\bA^k) + \nabla\hat u^k\|^2 + \|\div\hat\bA^k_t\|^2  \leq C ( \|\Theta\|^2_{H^1}+ \|\hat u^k_t \|^2 +  \|\nabla\hat u^k\|^2).\nonumber\\
\end{eqnarray}
Finally, multiplication in $L^2(\W)$ of (\ref{u_k}) by
$\sigma(\hat u_t^k + \hat u^k)$, with a (large enough) positive
constant $\sigma$ leads to
\begin{equation}\label{Gk}
    \frac{d}{dt}\left[c_0\sigma \|\hat u^k \|^2 + k_0\sigma \|\nabla\hat u^k\|^2 \right]+ c_0\sigma\|\hat u^k_t \|^2 + k_0\sigma\|\nabla\hat u^k\|^2 \leq C\ \|\Gamma\|^2.
\end{equation}
From the definition of $\Upsilon, \Theta, \Gamma$, interpolation inequality (\ref{H1H2}) and (\ref{comp}) it follows
$$
\|\Upsilon\|^2_{H^1}+\|\Theta\|^2_{H^1}+\|\Gamma\|^2\leq C_R(1+\|\psi_t\|^2).
$$
Therefore, in view of Remark \ref{remark}, we obtain
\begin{equation}\label{CR}
    \|\Upsilon\|^2_{H^1}+\|\Theta\|^2_{H^1}+\|\Gamma\|^2\leq C_R.
\end{equation}
Summing up (\ref{Uk})-(\ref{Gk}), with a properly choice of $\sigma$, on account of (\ref{CR}), we obtain
\begin{eqnarray}
&&
\frac{d}{dt} \cF^k+ \frac{1}{\k^2}\|\triangle\psi^k \|^2 +2\|\nabla\psi^k\|^2+ \mu \|\rot\rot\hat\bA^k\|^2
\nonumber \\
\label{FK}
&&
+ \|-\nabla(\div\hat\bA^k) + \nabla\hat u^k\|^2   + C\|\nabla\hat u^k\|^2
    \leq C_R,
\end{eqnarray}
where
\bea
\cF^k &=& \frac{1}{\k^2}\|\triangle\psi^k \|^2 + (\gamma + 1)\|\nabla\psi^k\|^2+ \mu \|\rot\rot\hat\bA^k\|^2 + \|\rot\hat\bA^k\|^2 
\\
&+&  \|\div\hat\bA^k\|^2+\|-\nabla(\div\hat\bA^k) + \nabla\hat u^k\|^2 +  c_0\sigma
\|\hat u^k \|^2 + k_0\sigma \|\nabla\hat u^k\|^2. \eea By adding
to both sides of (\ref{FK}) the terms
$\e(\|\rot\hat\bA^k\|^2+\|\div\hat\bA^k\|^2+\|\hat u^k\|^2)$, with
a small positive constant $\e$, from the Poincar\'e inequality we
prove
$$
\frac{d}{dt}\cF^k + \lambda \cF^k \leq C_R,
$$
where $\l>0$.
Owing to (\ref{bc_k}), an application of Gronwall's lemma yields
$$
\cF^k \leq C_R.
$$
Moreover, from (\ref{comp}) and (\ref{zc}) it follows that
$$
\|z^k(t)\|_{\cZ^1} \leq \|z(t)\|_{\cZ^1} + \|z^l(t)\|_{\cZ^1} \leq C_R,
$$
so that (\ref{zk}) is satisfied.

Relations (\ref{zc}) and (\ref{zk}) lead to
\begin{equation}\label{d1}
\text{dist}_{\cZ^1} (S(t)\cB, \tilde\cB_2) \leq m_1(R) e^{-\nu t}.
\end{equation}
Since $\tilde\cB_2$ is bounded in $\cZ^2(\W)$, we deduce that
$$
S(t)\tilde\cB_2\subset \cB_2,\qquad \forall t>\tilde{t}_2=\tilde{t}_2(R).
$$
Accordingly, we obtain
$$
{\rm dist}_{\cZ^1}(S(t)\tilde\cB_2, \cB_2)=
\begin{cases}
\a(R) & t\leq\tilde{t}_2
\\
0 & t>\tilde{t}_2
\end{cases}
$$
where $\a(R)$ is a bounded function. Hence, there exists an increasing function $m_3=m_3(R)$ such that
\begin{equation}\label{d2}
{\rm dist}_{\cZ^1}(S(t)\tilde\cB_2, \cB_2)\leq \a(R)e^{\tilde{t}_2-t}= m_3(R)e^{-t}.
\end{equation}
Inequalities (\ref{d1}), (\ref{d2}) and the transitivity property  of exponential attraction (see \cite[Theor 5.1]{FGMZ})
prove (\ref{expB2}).

Condition ({\it ii}) follows directly by the definition of absorbing set.
\qed

Now we are in a position to prove the main result of this section.
\begin{theor}
The semigroup $S(t)$ possesses an exponential attractor $\cE \subset \cZ^2(\W)$.
\end{theor}

\noindent
{\bf Proof.}
We apply lemma \ref{lemma1} with $\cK=\cB_2$ and $t^*>t_2$. Firstly we prove condition ({\it i}). Let $t^*\leq t\leq\tau\leq 2t^*$ and $z_1,z_2\in\cB_2$. Then we have
\begin{equation}\label{triang}
    \|S(\tau)z_1 -S(t)z_2\|_{\cZ^1} \leq \|S(\tau)z_1 -S(\tau)z_2\|_{\cZ^1} + \|S(\tau)z_2 -S(t)z_2\|_{\cZ^1}.
\end{equation}
Theorem \ref{cont_dep} implies
$$
\|S(\tau)z_1 -S(\tau)z_2\|_{\cZ^1} \leq C(t^*)\|z_1 -z_2\|_{\cZ^1}.
$$
In order to estimate last term in (\ref{triang}), we integrate (\ref{conti}) with respect to $t$. By taking (\ref{apriori_z3}) and (\ref{comp}) into account, we deduce
$$
\int_0^t \left[\|\nabla\psi_t\|^2+\|\rot\hat\bA_t\|^2+\|\div\hat\bA_t\|^2+\|\hat u_t\|^2\right] ds \leq C(t).
$$
The latter, together with a-priori estimate (\ref{apriori_z2}),
assures that
$$
\int_0^t \|z_t\|^2_{\cZ^1} ds \leq C(t).
$$
An application of H\"older's inequality leads to
$$
\|S(\tau)z_2 -S(t)z_2\|_{\cZ^1} \leq \int_t^\tau \|z_{2t}(s)\|_{\cZ^1} ds \leq C(t^*) \sqrt{\tau - t}.
$$
Hence, (\ref{triang}) reads
$$
\|S(\tau)z_1 -S(t)z_2\|_{\cZ^1} \leq C(t^*)\left[\|z_1 -z_2\|_{\cZ^1} + \sqrt{\tau - t}\right].
$$

Now we show that condition ({\it ii}) of lemma \ref{lemma1} holds.
We define
$$
L(z)=z^l(t^*),
$$
with $z^l(t)$ solution to (\ref{psi_l})-(\ref{bc_l}) with initial datum $z$. If $z_1^l(t), z_2^l(t)$ are solutions of (\ref{psi_l})-(\ref{bc_l}) with initial data $z_1$, $z_2$, their difference satisfies the same inequality (\ref{zc}), namely
$$
\|z_1^l(t)-z_2^l(t)\|_{\cZ^1}\leq C e^{-\nu t^*} \|z_1-z_2\|_{\cZ^1}.
$$
By choosing a sufficiently large $t^*>t_2$, we prove
$$
\|L(z_1)-L(z_2)\|_{\cZ^1} \leq \l \|z_1 -z_2\|_{\cZ^1},
$$
with $\l\in(0,1/2)$.

Let
$$
K(z)=z^k(t^*),
$$
with $z^k(t^*)$ solution to (\ref{psi_k})-(\ref{bc_k}). We denote
by $\psi^k=\psi_1^k-\psi_2^k$, $\psi=\psi_1-\psi_2$,
$\hat\bA^k=\hat{\bA}_1^k-\hat{\bA}_2^k$,
$\hat\bA=\hat{\bA}_1-\hat{\bA}_2$, $\hat
u^k=\hat{u}_1^k-\hat{u}_2^k$, $\hat u=\hat{u}_1-\hat{u}_2$. Let us
multiply \eqref{psi_k} by $1/2\,(\bar{\psi}_t+\triangle
\bar\psi_t)$, its conjugate by $1/2\,(\psi_t+\triangle \psi_t)$
and integrate over $\W$. Moreover, by proceeding like in the proof
of lemma \ref{lemma2} for the remaining equations \eqref{A_k} and
\eqref{u_k}, from (\ref{psi_k})--\eqref{bc_k} we deduce
\begin{eqnarray*}
&&\frac12 \frac{d}{dt}\bigg[\frac{1}{\k^2}\|\triangle\psi^k\|^2 + \left(\frac{1}{\k^2}+1\right)\|\nabla\psi^k\|^2+ \|\psi^k\|^2 +\|\nabla(\div \hat{\bA}^k)\|^2\\
&&+ \mu\|\rot\rot\hat{\bA}^k \|^2 +k_0\sigma\|\nabla\hat u^k\|^2\bigg]
    \\
    &&\leq
    C[\|\Upsilon_1-\Upsilon_2\|_{H^1} ^2+ \|\Theta_1-\Theta_2\|_{H^1}^2 +  \|\Gamma_1-\Gamma_2\|^2].
\end{eqnarray*}
where $\sigma>0$ is a suitable constant and
\begin{eqnarray*}
\Upsilon_1 -\Upsilon_2 &=&-\frac {2i}{\kappa} [\hat\bA\cdot \nabla \psi_1+(\hat\bA_2+\bA_{\cH})\cdot \nabla \psi]-|\hat\bA_1+\bA_{\cH}|^2 \psi 
 \\
  &&- \psi_2(\hat\bA_1+\hat\bA_2+2\bA_{\cH})\cdot\hat\bA+i\b(\psi\div\hat\bA_1+\psi_2\div\hat\bA)
  +2\psi\nonumber
  \\
  &&
  -\psi|\psi_1|^2 -\psi_2(\bar\psi_1\psi+\psi_2\bar\psi) -\psi
  (\hat{u}_1+u_{\cH})-\psi_2\hat{u}
\\
\Theta_1 - \Theta_2 &=& -|\psi_1|^2\hat{\bA}-(\bar\psi_1\psi+\psi_2\bar\psi)(\hat\bA_2+\bA_{\cH})\\
&&+\frac {i}{2\kappa} \big(\psi\nabla \bar\psi_1-\bar \psi \nabla \psi_1+\psi_2\nabla \bar\psi-\bar \psi_2 \nabla \psi\big)
\\
\G_1 -\G_2 &=& \frac{1}{2}(\psi_{1t}\bar\psi + \bar\psi_2 \psi_t + \bar\psi_{1t}\psi + \psi_2 \bar\psi_t)+
\div\Big[-|\psi_1|^2 \hat{\bA}  
 \\
 && -(\bar\psi_1\psi+\psi_2\bar\psi)
(\hat {\bA}_2+\bA_{\cH}) +
  \frac {i}{2\kappa} \big(\psi\nabla \bar\psi_1-\bar \psi \nabla \psi_1+\psi_2\nabla \bar\psi-\bar \psi_2 \nabla \psi\big)\Big]
\end{eqnarray*}

Therefore, since $z_1, z_2\in\cB_2$ in view of (\ref{H1H2}), we easily deduce
$$
\left\|\Upsilon_1 - \Upsilon_2\right\|_{H^1}+\left\|\Theta_1 - \Theta_2\right\|_{H^1} +\left\|\Gamma_1 - \Gamma_2\right\|
 \leq C[\|\psi\|_{H^2} + \|\hat {\bA}\|_{\cH^2} +  \|\hat u\|_{H^1_0}].
$$

Hence
\begin{eqnarray*}
&&\frac12 \frac{d}{dt}\bigg[\frac{1}{\k^2}\|\triangle\psi^k\|^2 +
\left(\frac{1}{\k^2}+1\right)\|\nabla\psi^k\|^2+ \|\psi^k\|^2
+\|\nabla(\div \hat{\bA}^k)\|^2\\
&&+ \mu\|\rot\rot\hat{\bA}^k \|^2
+k_0\sigma\|\nabla\hat u^k\|^2\bigg]
    \\
    &&\leq
    C[\|\psi\|_{H^2} ^2+ \|\hat {\bA}\|_{\cH^2}^2 +  \|\hat u\|_{H^1_0}^2],
\end{eqnarray*}
which in view of (\ref{int_Z2}) implies
$$
\|K(z_1,z_2)\|_{\cZ^2} \leq \L \|z_1 - z_2\|_{\cZ^1},
$$
with a suitable constant $\L>0$.

Assumptions $(i)$ and $(ii)$ of lemma \ref{lemma1} hold, so that inequality (\ref{exp}) is satisfied.
The inequality (\ref{exp_rate}) follows by applying lemma \ref{lemma2} and the transitivity property  of exponential attraction.
\qed

\bigskip

\noindent
{\bf Acknowledgement.}
The authors have been partially supported by G.N.F.M. - I.N.D.A.M. through the project for young researchers ``Phase-field models for second-order transitions''.

\end{document}